\documentclass[journal]{IEEEtran}
\def\BibTeX{{\rm B\kern-.05em{\sc i\kern-.025em b}\kern-.08em T\kern-.1667em\lower.7ex\hbox{E}\kern-.125emX}}

\usepackage{amsmath,amssymb,amsfonts}
\usepackage{algorithmic}
\usepackage{array}
\usepackage{graphicx}
\usepackage{textcomp}
\usepackage{url}
\usepackage{verbatim}
\usepackage{balance}
\usepackage{silence}
\usepackage[T1]{fontenc}               
\usepackage[utf8]{inputenc}            
\usepackage{amsmath,amssymb,amsfonts}  
\usepackage{newtxtext,newtxmath}       
\usepackage{mathrsfs}                  
\usepackage{anyfontsize}               
\usepackage{microtype}                 
\usepackage{pifont}
\usepackage[
  backend=biber,
  style=ieee,
  sorting=none,
  doi=false,
  isbn=false,
  url=false,
  eprint=false
]{biblatex}

\usepackage{graphicx}                  
\usepackage{siunitx}                   
\usepackage{tabularray}                
\UseTblrLibrary{diagbox}
\UseTblrLibrary{booktabs}              
\usepackage{diagbox} 

\usepackage[caption=false,font=normalsize,labelfont=sf,textfont=sf]{subfig} 
\usepackage{stfloats}                  

\usepackage{enumitem}                  
\usepackage{xparse}                    
\usepackage{etoolbox}                  

\usepackage[dvipsnames,table]{xcolor}  
\usepackage{pgf,tikz,pgfplots}         
\pgfplotsset{compat=1.18}              
\usetikzlibrary{%
  arrows.meta,arrows,                  
  backgrounds,fit,positioning,         
  shapes,shapes.geometric,             
  calc,                                
  decorations.pathmorphing             
}

\definecolor{orange7}{RGB}{255,100,0}  
\usepackage{ninecolors}                

\usepackage[export]{adjustbox}         
\usepackage[skins,listings,theorems]{tcolorbox} 
\tcbset{contrastbox/.style={
    colback=gray!5, colframe=black!40,
    boxrule=0.3pt, arc=2pt, left=4pt, right=4pt, top=2pt, bottom=2pt,
    fonttitle=\bfseries, coltitle=black, titlerule=0pt,
}}

\usepackage{float}                     
\usepackage{standalone} 
\usepackage{placeins}
\newtheorem{remark}{Remark}

\usepackage{orcidlink}                 
\newcommand{\fatpill}{10pt}
\newcommand{\longpill}{10pt}

\newcommand{\blocktextwidth}{4.6cm}

\newcommand{\tightitems}{%
    \setlength{\itemsep}{0pt}%
    \setlength{\parskip}{0pt}%
    \setlength{\parsep}{0pt}%
    \setlength{\topsep}{0pt}%
    \setlength{\partopsep}{0pt}%
}
\newcommand{\epm}{EPM}
\newcommand{\dpm}{DPM}
\newcommand{\ddm}{DDM}

\newcommand{\pidc}{PID}
\newcommand{\risec}{RISE}
\newcommand{\nmpcc}{N-NMPC}
\newcommand{\anmpc}{A-NMPC}
\newcommand{\rnmpc}{R-NMPC}

\newcommand{\norm}[1]{\left\lVert #1 \right\rVert}

\NewDocumentCommand{\rc}{s m}{%
  \IfBooleanTF{#1}%
    {{\color{black}#2}}%
    {{\color{red}#2}}%
}

\NewDocumentCommand{\bc}{s m}{%
  \IfBooleanTF{#1}%
    {{\color{black}#2}}%
    {{\color{blue}#2}}%
}

\NewDocumentCommand{\gc}{s m}{%
  \IfBooleanTF{#1}%
    {{\color{black}#2}}%
    {{\color{magenta}#2}}%
}

\definecolor{pidcolor}{HTML}{F0C571}      
\definecolor{risecolor}{HTML}{59A89C}     
\definecolor{nmpccolor}{HTML}{4A559A}     
\definecolor{robustcolor}{HTML}{CCB9B9}   
\definecolor{adaptivecolor}{HTML}{082A54} 
\definecolor{DPM}{HTML}{97a6c4}
\definecolor{DDM}{HTML}{5f7396}
\definecolor{EPM}{HTML}{384860}

\begin{document}

\title{CAPE: Control Algorithm Performance Evaluation under Learned Vehicle Dynamics Models}

\author{Malik Ali$^*$ \IEEEmembership{Student Member, IEEE},
Musabbir Ahmed Arrafi$^+$ \IEEEmembership{Student Member, IEEE},\\
Nicholas M. Stiffler \IEEEmembership{Member, IEEE},
and Krishna Bhavithavya Kidambi \IEEEmembership{Senior Member, IEEE}%
\thanks{The authors are with the University of Dayton, Dayton, OH 45469 USA.
M. Ali is with Electrical and Computer Engineering;
M. A. Arrafi and K. B. Kidambi are with Mechanical and Aerospace Engineering;
and N. M. Stiffler is with Computer Science.
E-mail: alim37@udayton.edu, arrafim1@udayton.edu, nstiffler1@udayton.edu, kkidambi1@udayton.edu.}
\thanks{$^*$M. Ali and $^+$M. A. Arrafi contributed equally to this work.}}


\maketitle

\begin{abstract}
%
%

We propose the Control Algorithm Performance Evaluation (CAPE) framework, a systematic methodology for benchmarking racing controllers under our proposed learned enhanced physics model (\epm{}). 
The proposed framework enables cross-controller comparison by evaluating five closed-loop control architectures.
We further compare our proposed \epm{} with two state-of-the-art learned vehicle dynamics models: Deep Pacejka Model (\dpm{}) and Deep-learning Dynamics Model (\ddm{}).
\bc*{Closed-loop experiments show that across all models and controllers, the proposed \epm{} achieves best average lap times. Specifically, the Adaptive NMPC with \epm{} achieves a time of $5.82\,\si{\second}$, compared with $12.99\,\si{\second}$ for \dpm{} and $8.80\,\si{\second}$ for \ddm{}, while simultaneously producing substantially lower longitudinal and lateral tracking errors under identical controller configurations.}

We further evaluate all three models and five controllers using a disturbance-aware simulation framework incorporating measurement noise, process disturbances, actuator delay, and parametric uncertainty. \bc*{Under moderate global disturbance scaling factor ($\eta = 1$), results averaged across the five controllers show that \epm{} reduces a) longitudinal tracking error by $29.0\%$ and $17.2\%$; b) lateral tracking error by $24.6\%$ and $12.3\%$; c) while increasing average velocity magnitude by $39.9\%$ and $3.1\%$ relative to \dpm{} and \ddm{}, respectively.}
Overall, CAPE establishes a systematic benchmark for evaluating the performance of learned vehicle dynamics models in a closed-loop control framework and demonstrates that our proposed \epm{} significantly improves controller robustness and performance under realistic uncertainties.
%
%
%
%
\end{abstract}


\section{Introduction}

Autonomous racing provides a demanding testbed for evaluating control algorithms that operate near the limits of vehicle dynamics \cite{moon2024autonomous,thakkar2024hierarchical}. 
At racing speeds, controllers must handle rapid acceleration, aggressive cornering, and nonlinear tire behavior where small modeling errors can destabilize closed-loop performance \cite{albhaisi2026trajectory,betz2022autonomous}. 
%
%
As a result, stand-alone controller evaluation in autonomous racing cannot achieve optimal results without accurate open-loop vehicle modeling. 
%
%
Therefore, closed-loop performance evaluation in racing requires a benchmarking framework that accounts for both vehicle model and control architecture in tandem. 

Classical physics-based vehicle dynamic models such as bicycle model or multi-body formulations provide computationally efficient approximations. 
%
However, these simplified models diverge significantly from real vehicle behavior at high accelerations and near the tire friction limits, where accurate tire modeling and parameterization become critical for reliable closed-loop control performance \cite{zhang2024survey1,musiu2026comprehensive}.
Recent research has explored data-driven and hybrid modeling approaches that learn vehicle dynamics directly from data, including models such as the Deep Pacejka Model (\dpm{}) \cite{kim2022dpm} and Deep Dynamics Model (\ddm{}) \cite{chrosniak2024deep}, which aim to improve predictive accuracy while remaining compatible with control frameworks.
\dpm{} \cite{kim2022dpm} embeds the Pacejka tire model as a layer within the network, allowing it to learn tire force parameters while preserving the structure of the vehicle bicycle dynamics model.
The authors in \ddm{} \cite{chrosniak2024deep} embed a neural network within a physics-constrained single-track vehicle model to capture nonlinear tire and drivetrain dynamics from data.
%
%
A hyperparameter optimization-based identification method that efficiently optimizes the tire and engine models for full-scale autonomous racing vehicles and integrates the resulting models into planning and control systems is proposed in \cite{seong2023model}.
The authors in \cite{sonmez2024symmetry} exploit symmetry in system dynamics to learn lower-dimensional models that improve accuracy and sample efficiency in model-based reinforcement learning.
While these approaches improve predictive accuracy, their robustness under model mismatch, sensing uncertainties, and actuator delays is insufficiently studied, and systematic benchmarks evaluating their impact on closed-loop control performance are still lacking.

In parallel, a plethora of control algorithms have been explored in the context of autonomous racing including linear \pidc{}, back-stepping \cite{selman2025lateral} geometric pursuit controllers \cite{bousskoul2025control}.
Optimization-based controllers such as nonlinear model predictive control remain prominent because they explicitly encode vehicle dynamics, constraints, and preview information over a receding horizon, enabling operation close to the handling limits \cite{kabzan2019learning,joa2024piecewise}. 
The authors in \cite{dai2024mpc} design a model predictive controller (MPC) for a nominal dynamics model along with past runtime data and use Gaussian process regression to capture the mismatch between the nominal and true model. 
%
Uncertainty-aware trajectory planning approaches extend MPC with a trajectory prediction method combining kinematics and reachable set in order to consider the uncertain trajectory of the target vehicle to enhance safety \cite{guo2025uncertainty}.
Even though model-based controller approaches provide an effective framework for trajectory tracking and constraint handling, ensuring safety and stability when incorporating learned components remains a critical challenge \cite{liu2025regret}. 

Consequently, recent work has focused on developing certificate-based approaches, where neural controllers are trained or verified together with Lyapunov, barrier, or contraction certificates \cite{dawson2023survey}.
The authors in \cite{wang2024actorcritic} propose an algorithmic framework for co-learning a neural network control policy and a neural network Lyapunov function for actuation-constrained nonlinear systems. 
A method to train neural network controllers with guaranteed disk-type stability margins is proposed in \cite{junnarkar2025stability}, under uncertainties and nonlinearities.  
Together, these works demonstrate that learning-based controllers provide formal stability guarantees. However, these approaches are typically evaluated within a fixed controller–model pairing, limiting insight into how learned dynamics models affect stability and performance across heterogeneous control architectures.

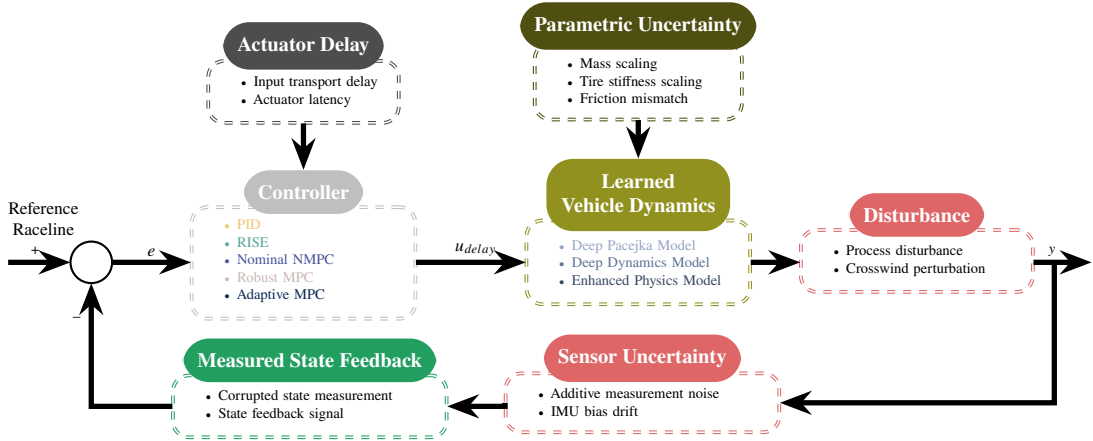
\begin{figure*}[b]
\centering
\begin{tikzpicture}[
    scale=0.55,
    transform shape,
    >={Stealth},
    signal/.style={->, line width=0.7mm},
    title/.style={
        rounded corners=10pt,
        fill,
        text=white,
        inner xsep=\longpill,
        inner ysep=\fatpill,
        font=\Large\bfseries
    },
    dashedblock/.style={
        dashed,
        double,
        rounded corners=6pt,
        inner sep=6pt
    },
    blockbody/.style={
        align=left,
        text width=\blocktextwidth,
        inner sep=0pt,
        font=\normalsize
    }
]

\node[
    blockbody,
    anchor=north west
] (controllertext) at (-10.55,2.00) {
    \vspace*{-1em}
    \begin{itemize}
        \tightitems
        {\color{pidcolor} \item PID}
        {\color{risecolor} \item RISE}
        {\color{nmpccolor} \item Nominal NMPC}
        {\color{robustcolor} \item Robust MPC}
        {\color{adaptivecolor} \item Adaptive MPC}
    \end{itemize}
};

\node[
    dashedblock,
    draw=gray8,
    fit=(controllertext),
    minimum width=2cm
] (controller) {};

\node[fill=gray8, title] (controllertitle) at ($(controller.north)+(0,0.35)$) {Controller};

\node[
    blockbody,
    anchor=north west
] (modeltext) at (-2.50,1.68) {
    \vspace*{-0.5em}
    \begin{itemize}
        \tightitems
        {\color{DPM} \item Deep Pacejka Model}
        {\color{DDM} \item Deep Dynamics Model}
        {\color{EPM} \item Enhanced Physics Model}
    \end{itemize}
};

\node[
    dashedblock,
    draw=yellow6,
    fit=(modeltext)
] (model) {};

\node[fill=yellow6, title, align=center] (modeltitle) at ($(model.north)+(0,0.60)$) {Learned \\ Vehicle Dynamics};

\node[
    blockbody,
    anchor=south,
    text width=3.8cm
] (delaytext) at ($(controller.north)+(0,2.4)$) {
    \vspace*{-0.8em}
    \begin{itemize}
        \tightitems
        \item Input transport delay
        \item Actuator latency
    \end{itemize}
};

\node[
    dashedblock,
    draw=gray3,
    fit=(delaytext)
] (delaybox) {};

\node[fill=gray3, title] at ($(delaybox.north)+(0,0.35)$) {Actuator Delay};

\node[
    blockbody,
    anchor=south,
    text width=4.2cm
] (paramtext) at ($(model.north)+(0,2.8)$) {
    \vspace*{-0.8em}
    \begin{itemize}
        \tightitems
        \item Mass scaling
        \item Tire stiffness scaling
        \item Friction mismatch
    \end{itemize}
};

\node[
    dashedblock,
    draw=yellow3,
    fit=(paramtext)
] (parambox) {};

\node[fill=yellow3, title, align=center] at ($(parambox.north)+(0,0.40)$) {Parametric Uncertainty};

\node[
    blockbody,
    anchor=north west,
    text width=4.8cm
] (disttext) at (4.10,1.45) {
    \vspace*{-0.8em}
    \begin{itemize}
        \tightitems
        \item Process disturbance
        \item Crosswind perturbation
    \end{itemize}
};

\node[
    dashedblock,
    draw=red6,
    fit=(disttext)
] (distbox) {};

\node[fill=red6, title, align=center] at ($(distbox.north)+(0,0.35)$) {Disturbance};

\node[
    blockbody,
    anchor=north west,
    text width=5.8cm
] (meastext) at (-3.00,-2.00) {
    \vspace*{-0.8em}
    \begin{itemize}
        \tightitems
        \item Additive measurement noise
        \item IMU bias drift
    \end{itemize}
};

\node[
    dashedblock,
    draw=red6,
    fit=(meastext)
] (measbox) {};

\node[fill=red6, title, align=center] at ($(measbox.north)+(0,0.35)$) {Sensor Uncertainty};

\node[
    blockbody,
    anchor=north west,
    text width=5.8cm
] (sensetext) at (-11.00,-2.05) {
    \vspace*{-0.8em}
    \begin{itemize}
        \tightitems
        \item Corrupted state measurement
        \item State feedback signal
    \end{itemize}
};

\node[
    dashedblock,
    draw=teal6,
    fit=(sensetext)
] (sensebox) {};

\node[fill=teal6, title, align=center] at ($(sensebox.north)+(0,0.35)$) {Measured State Feedback};

\coordinate (sumcenter) at ($(controller.west |- model.center)+(-2.4,0)$);

\coordinate (cin)  at (controller.west |- model.center);
\coordinate (cout) at (controller.east |- model.center);

\coordinate (min)  at (model.west |- cout);
\coordinate (mout) at (model.east);

\coordinate (distin)  at (distbox.west |- mout);
\coordinate (distout) at (distbox.east |- distin);

\coordinate (yout)    at ($(distout)+(1.4,0)$);
\coordinate (yreturn) at ($(distout)+(0.5,0)$);

\coordinate (fbtap)   at ($(sumcenter)+(0,-0.5)$);

\draw[thick] (sumcenter) circle[radius=0.5];

\node at ($(sumcenter)+(-1.35,0.28)$) {$+$};
\node at ($(sumcenter)+(-0.35,-1.35)$) {$-$};

\node[align=center, font=\large] at ($(sumcenter)+(-1.15,1.05)$) {Reference \\ Raceline};

\draw[signal] ($(sumcenter)+(-2,0)$) -- ($(sumcenter)+(-0.5,0)$);

\draw[signal] ($(sumcenter)+(0.5,0)$) -- (cin)
    node[font=\large, midway, above] {$e$};

\draw[signal] (cout) -- (min)
    node[font=\large, pos=0.30, above right] {$u_{delay}$};

\draw[signal] (mout) -- (distin);

\draw[signal] (distout) -- (yout)
    node[pos=0.82, above, xshift=-20pt] {$y$};

\draw[signal] (delaybox.south) -- (controllertitle.north);
\draw[signal] (parambox.south) -- (modeltitle.north);

\draw[signal] (yreturn) |- (measbox.east);
\draw[signal] (measbox.west) |- (sensebox.east);
\draw[signal] (sensebox.west) -| (fbtap);

\end{tikzpicture}
\caption{Proposed closed-loop benchmarking architecture. The controller generates commands that pass through actuator delay before interacting with the learned dynamics model. Noises, disturbances, and uncertainty affect system evolution and corrupt the feedback signal used to compute the tracking error.}
\label{fig:blockdiagram}
\end{figure*}
Despite extensive research on both learned vehicle modeling and controllers for autonomous racing, the effect of learned dynamics model choice on heterogeneous closed-loop controller performance has not been systematically evaluated.
Controller benchmarking studies explain which control law performs better under a fixed internal model, while learned-model studies explain how to improve a particular dynamics representation for prediction or for one learning-enabled control framework \cite{kabzan2019learning,chrosniak2024deep,fang2024fine, charles2025advancing}. 
%

To address this problem, we propose the Control Algorithm Performance Evaluation (CAPE) framework, a systematic methodology for benchmarking autonomous racing controllers under learned vehicle dynamics models.
Within this framework, we compare our proposed \epm{} against \dpm{} and \ddm{} baselines across five representative control architectures: \pidc{} control, robust inverse of signum error (RISE) control, \bc*{nominal} nonlinear model predictive control (\nmpcc{}), robust NMPC (\rnmpc{}), and adaptive NMPC (\anmpc{}). 
By evaluating these heterogeneous controllers within a common simulation and reference-generation pipeline, CAPE isolates the effect of the dynamics on closed-loop control performance.
\bc*{
%
Classical bicycle dynamic models for racing have been studied in literature \cite{liniger_rc}, and their performance degrades near handling limits due to fixed parameters, transient model mismatches, tire saturation, load transfer and actuation delay. 
CAPE therefore focuses on isolating how learned dynamics affect closed-loop controller behavior under a common benchmark.}

The CAPE benchmark evaluates controller behavior using autonomous racing telemetry from a simulated autonomous racing platform with training and testing performed on separate racetracks to assess model generalization. 
\bc*{To emulate realistic autonomous racing conditions, the evaluation pipeline introduces sensing, actuation, and modeling uncertainties, including multiplicative measurement noise, process disturbances, actuator delays, IMU bias drift, crosswind perturbations, and parametric model mismatch, controlled through a global disturbance scaling factor ($\eta$). 
Fig~\ref{fig:blockdiagram} shows the overall proposed closed-loop benchmarking architecture.}
%
Controller performance is evaluated using longitudinal and lateral tracking accuracy, control input effort, and \bc*{one complete lap time} across the multiple controllers.
This setup enables direct comparison of how learned vehicle dynamics influence controller behavior under increasing levels of uncertainty.
The main contributions of this work are summarized as:
\begin{enumerate}
\item A unified benchmarking framework for evaluating heterogeneous autonomous racing controllers under three different learned models (\dpm{}, \ddm{}, \epm{}- proposed)
\item A disturbance-aware evaluation pipeline that systematically injects sensing, actuation, and modeling uncertainties to study controller robustness.
\item An empirical comparison of three learned vehicle dynamics models (DPM, DDM, and EPM) demonstrating how model structure influences tracking accuracy, control effort, and stability across multiple closed-loop controllers.
\end{enumerate}

The remainder of this paper is organized as follows. 
Section~\ref{sec:mathematical_model} presents the proposed learning-enhanced physics-aware vehicle dynamics model, along with \dpm{} and \ddm{}.  
Section~\ref{sec:network} describes the neural network architecture used to estimate the model parameters. 
Section~\ref{sec:closed_loop_control} introduces the five controller architectures. 
%
Section~\ref{benchmark} discusses the disturbance-aware benchmarking setup used in the CAPE framework. 
Section~\ref{sec:simulation_results} presents the closed-loop simulation results and comparative robustness analysis, and 
Section~\ref{sec:conclusion} concludes the paper. 

\bc*{\textbf{Dislaimer}: A preliminary version of this work is currently under review at IEEE Conference on Decision and Control, 2026. This paper substantially extends that version by adding open-loop learned-model evaluation, fixed-parameter bicycle-model comparisons, failure-penalized tracking, expanded uncertainty-sweep results, and additional discussion of MPC ablation study that isolates the effect of the internal controller model from the learned dynamics model.}
 
\section{Mathematical Model}
\label{sec:mathematical_model}
This section presents the control-oriented formulation of the learning-enhanced physics-aware vehicle dynamics model used in this study. 
The model combines a dynamic single-track representation with additional equations that include load-sensitive tire forces, longitudinal load transfer, and learned actuator behavior. 
%
\vspace{-0.5\baselineskip}

\subsection{Single-Track Vehicle Dynamics}
We consider a planar single-track vehicle model with state
\begin{equation}
\mathbf{x}_t =
\begin{bmatrix}
v_x & v_y & \omega
\end{bmatrix}^{\top} \in \mathbb{R}^3,
\end{equation}
where $v_x$ and $v_y$ denote longitudinal and lateral body-frame
velocities, and $\omega$ is the yaw rate. The control input is
\begin{equation}
\mathbf{u}_t =
\begin{bmatrix}
T & \delta
\end{bmatrix}^{\top} \in \mathbb{R}^2,
\end{equation}
where $T$ is the throttle command and $\delta$ is the steering angle.
The continuous-time dynamics are \cite{jazar2025vehicle}
\begin{subequations}
\label{dynamics}
\begin{align}
\dot{v}_x &= \frac{1}{m}\left(F_{rx} - F_{fy}\sin\delta + m v_y \omega \right),  \label{dyna1}\\
\dot{v}_y &= \frac{1}{m}\left(F_{fy}\cos\delta + F_{ry} - m v_x \omega \right), \\
\dot{\omega} &= \frac{1}{I_z}\left(l_f F_{fy}\cos\delta - l_r F_{ry}\right), \label{dyna3}
\end{align}
\end{subequations}
where $m$ is the vehicle mass, $I_z$ the yaw moment of inertia, and $l_f$, $l_r$ denote the distances from the center of gravity to the front and rear axles. 
$F_{rx}$ is the longitudinal force on rear tires, $F_{fy}, F_{ry}$ are lateral tire forces and $F_{fz}, F_{rz}$ are the normal forces on front and rear tires respectively. 
Considering the learned dynamic state in Eq.~(1), closed-loop tracking also requires the planar position and heading $[x, y, \psi]^T$. Therefore, standard bicycle-model kinematics are used ~\cite{liniger_rc}:
\begin{subequations}
\begin{align}
\dot{x} &= v_x\cos\psi - v_y\sin\psi, \label{dyna4} \\ 
\dot{y}  &= v_x\sin\psi + v_y\cos\psi,  \\
\dot{\psi} &= \omega. \label{dyna6}
\end{align}
\end{subequations}
The complete propagated state used by the controllers are $[x,y,\psi,v_x,v_y,\omega]^\top$.
\vspace{-0.75\baselineskip}
\subsection{Enhanced Physics Model (EPM)}
 In addition to the dynamics defined in Eq.~(\ref{dyna1})--(\ref{dyna3}), \epm{} embeds a) slip-angle formulations, b) dynamic load transfer, c) load-sensitive tire forces and d) actuator dynamics. 
Each training sample contains a temporal window of vehicle dynamic states and actuator command/feedback histories, and the supervised target to predict the next-step dynamic state $[v_x,v_y,\omega]^\top$. Readers are referred to \cite{arrafilepavd} for additional details on EPM.
\vspace{-0.75\baselineskip}
\subsection{ Deep Pacejka Model (DPM)}
 \dpm{}~\cite{kim2022dpm} retains a bicycle-model structure and embeds a Pacejka tire-force layer inside the learning architecture, allowing the network to learn tire-force behavior while preserving physically motivated vehicle-dynamics relationships. 
\vspace{-0.5\baselineskip}
\subsection{Deep Dynamics Model (DDM)}
 \bc*{\ddm{}~\cite{chrosniak2024deep} uses a Physics-Constrained Neural Network to estimate Pacejka coefficients and drivetrain resistances within a single-track model. While effective, it assumes fixed tire parameters and neglects load-dependent tire
forces, nominal tire-force variation, suspension effects, and actuator delays.}

\section{Network Architecture}
\label{sec:network}
The proposed Enhanced Physics Model (\epm{}) employs a compact neural
parameter estimator to learn physically interpretable coefficients
appearing in the vehicle dynamics formulation. Instead of directly
predicting future states, the network estimates a vector of model
parameters that govern tire forces, drivetrain behavior, and actuator
dynamics.

Let $\Phi$ denote the vector of parameters appearing in the dynamics equations (\ref{dyna1})-(\ref{dyna3}). The network processes a short temporal window of vehicle states and control inputs,
\begin{equation}
Z_t = \{x_{t-k:t},\, u_{t-k:t}\},
\end{equation}
using gated recurrent units (GRU) to capture temporal dependencies in vehicle response,
\begin{equation}
h_t = \mathrm{GRU}(Z_t)
\end{equation}

The resulting feature vector is passed through a multilayer perceptron that predicts the parameter vector
\begin{equation}
\Phi = f_\theta(h_t),
\end{equation}
which includes tire model coefficients, drivetrain parameters, and actuator-related variables. To ensure physically
meaningful predictions, the outputs are constrained within predefined bounds using a physics guard layer.
The predicted parameters are inserted into the \epm{} vehicle dynamics equations to compute state evolution,
\begin{equation}
\dot{\textbf{x}} = f(\textbf{x},u,\Phi)
\end{equation}
This hybrid architecture combines data-driven parameter estimation with a physics-based vehicle model, enabling accurate and computationally efficient dynamics prediction for closed-loop control.
Table~\ref{tab:model_complexity} compares the architectural complexity and inference efficiency of the learned dynamics models. \epm{} achieves a substantial reduction in parameter count and floating-point operations compared to \ddm{}, while also attaining the lowest inference latency, demonstrating improved performance and efficiency for real-time closed-loop deployment. In addition, despite \epm{} having $30 \times$ more parameters than \dpm{}, \epm{} achieves a $23\%$ reduction in inference time. 

\begin{table}[htbp]
\centering
\caption{Model Complexity and Inference Efficiency}
\label{tab:model_complexity}

\resizebox{\columnwidth}{!}{
\begin{tblr}{
colspec={|l|c|c|c|c|c|c|},
row{1}={font=\bfseries},
hline{1-5}={-}{},
hline{2}={-}{},
row{2} = {bg=DPM!60},
row{3} = {bg=DDM!60},
row{4} = {bg=EPM!60},
}
Model & {Total\\Params (M)} & Size (MB) & {GRU \\layers} & {FC \\layers} & FLOPs (M) & {Avg. Inf.\\Time (ms)} \\
DPM & 0.028 & 0.11 & 8 & 3 & 0.023 & 0.081 \\
DDM & 2.878 & 10.98 & 7 & 17 & 5.740 & 0.072 \\
EPM & 0.917 & 3.46 & 7 & 8 & 1.802 & 0.063 \\
\end{tblr}
}
\end{table}
\begin{remark}
\bc*{The complete architecture and training details for \dpm{}, \ddm{}, and \epm{} are provided in~\cite{kim2022dpm}, ~\cite{chrosniak2024deep} and ~\cite{arrafilepavd} respectively. 
The expected closed-loop advantage of \epm{} comes from combining a constrained physics-based parameterization with a lower-complexity recurrent estimator. 
}
\end{remark}
\begin{remark}
\bc*{All 3 learned models considered for benchmarking differ in neural architecture, underlying dynamics formulation, and training objective. However, all 3 models are trained using the same data~\cite{jain2021bayesrace} with exact data split, input signals, and evaluated under identical closed-loop benchmarking protocol to ensure a meaningful performance comparison.}
\end{remark}
\vspace{-0.3cm}

\section{Closed-Loop Control}
\label{sec:closed_loop_control}
\bc*{To evaluate the influence of dynamics model fidelity on controller performance, we conduct closed-loop simulations using multiple control architectures with each controller interacting with the same vehicle dynamics interface described in Section~\ref{sec:mathematical_model}.}
The control objective is to regulate all 3 learned-models to track a predefined sufficiently smooth, time-varying reference trajectory, despite the presence of structured and unstructured uncertainties in the dynamic model, measurement noise, disturbances, and actuator delays.
Thus, the control objective can be mathematically stated as
\begin{equation} \label{clerror}
   \vert e \vert \rightarrow 0
\end{equation}

where the error $e = [e_{te}, e_{\psi}, e_{vte}]^T \in \mathbb{R}^5$ is defined as 
\begin{eqnarray} 
    e_{te} &=& [e_x, e_y]^T  \notag \\
            &=& [x_{ref}-x, y_{ref}-y]^T \label{ete} \\
    e_{\psi} &=& [\psi_{ref} - \psi] \label{epsi} \\
    e_{vte} &=& [e_{V_x}, e_{V_y}]^T \notag \\ 
            &=& [V_{x_{ref}}-V_x, V_{y_{ref}}-V_y]^T \label{evte}
\end{eqnarray} 
where $e_{te}$ is the tracking error, $e_{\psi}$ is the heading error and $e_{vte}$ is the velocity tracking error. 
The reference racing line is generated using the procedure outlined in \cite{jain2021bayesrace}. 
%
\bc*{At every step, the controller receives the current vehicle state and desired reference information and produces steering and throttle commands. 
These steering and throttle commands are then used to propagate the closed-loop vehicle states using the selected learned dynamics model \dpm{} \cite{kim2022dpm}, \ddm{} \cite{chrosniak2024deep}, and the proposed enhanced physics model (\epm{}).
MPC-based controllers use their respective learned models inside the optimizer which again propagates the state using the learned dynamics model. 
In addition, we show ablation studies to isolate the effect of the internal controller model from the
learned dynamics model used for closed-loop propagation.}
The controller settings used in the closed-loop benchmark are summarized in Table~\ref{tab:controller_params}. For each controller, these parameters are held fixed when switching between \dpm{}, \ddm{}, and \epm{} so that performance differences are not caused by controller retuning.
\begin{remark}
    \bc*{The focus of this paper is not to provide stability proofs for each controller. Instead, we evaluate how learned dynamics model choices affect tracking accuracy, robustness, and lap-completion time under various controller configurations in a unified environment.} 
\end{remark}
\begin{table}[ht]
\centering
\caption{CAPE: Closed-loop Controller Parameters}
\label{tab:controller_params}

\begin{tblr}{
    row{1} = {font=\bfseries},
    row{2} = {bg=pidcolor!60,fg=black,font=\bfseries\scriptsize},
    row{3-6} = {font=\scriptsize},
    colspec = {X[0.26,c,m] X[0.5,c,m] X[0.24,c,m]},
    colsep = 1mm,
    rowsep = .2mm,
    vlines = {1pt, black},
    hlines = {1pt, black},
    hline{4-5} = {dashed,fg=gray},
}
Parameter & Description & Value \\ \hline
\SetCell[c=3]{c} PID & & \\
$K_p^{e_{V_x}},K_i^{e_{V_x}},K_d^{e_{V_x}}$ & Velocity tracking-error gains & $1.20,\;0.30,\;0.02$ \\
$K_p^\psi,K_i^\psi,K_d^\psi$ & Heading-error gains & $2.20,\;0.05,\;0.08$ \\
$K_p^{cte},K_i^{cte},K_d^{cte}$ & Tracking-error gains & $1.00,\;0.02,\;0.05$ \\
\end{tblr}

\begin{tblr}{
    row{1} = {bg=risecolor!60,fg=black,font=\bfseries\scriptsize},
    row{3-9} = {font=\scriptsize},
    colspec = {X[0.26,c,m] X[0.5,c,m] X[0.24,c,m]},
    colsep = 1mm,
    rowsep = .2mm,
    vlines = {1pt, black},
    hlines = {1pt, black},
    hline{3-5} = {dashed,fg=gray},
}
\SetCell[c=3]{c} RISE & & \\
$k_{e_{V_x}}$ & Velocity tracking-error gain & $3.0$ \\
$k_u$ & Robust sign-feedback gain & $0.25$ \\
$k_s$ & Tracking-error feedback gain & $1.8$ \\
$\beta$ & Integral sign-error gain & $0.55$ \\
\end{tblr}

\begin{tblr}{
    row{1} = {bg=nmpccolor!60,fg=black,font=\bfseries\scriptsize},
    row{3-8} = {font=\scriptsize},
    colspec = {X[0.26,c,m] X[0.5,c,m] X[0.24,c,m]},
    colsep = 1mm,
    rowsep = .2mm,
    vlines = {1pt, black},
    hlines = {1pt, black},
    hline{3-7} = {dashed,fg=gray},
}
\SetCell[c=3]{c} Nominal NMPC & & \\
$N$ & Prediction horizon & $15$ \\
$T_s$ & Sampling time & $0.02$ s \\
$Q$ & Tracking weight & $\mathrm{diag}(1,1)$ \\
$P$ & Terminal tracking weight & $\mathrm{diag}(0,0)$ \\
$R$ & Input-rate weight & $\mathrm{diag}(0.005,1)$ \\
Solver & Nonlinear program solver & CasADi/IPOPT, $100$ max iter. \\
\end{tblr}

\begin{tblr}{
    row{1} = {bg=adaptivecolor!45,fg=black,font=\bfseries\scriptsize},
    row{3-7} = {font=\scriptsize},
    colspec = {X[0.26,c,m] X[0.5,c,m] X[0.24,c,m]},
    colsep = 1mm,
    rowsep = .2mm,
    vlines = {1pt, black},
    hlines = {1pt, black},
    hline{3-6} = {dashed,fg=gray},
}
\SetCell[c=3]{c} Adaptive NMPC & & \\
$\Theta$ & Online-updated parameters & $C_{m1},C_{r0},C_{r2}$ \\
$n_{\mathrm{upd}}$ & Parameter update interval & $5$ steps \\
$\alpha_{\mathrm{ad}}$ & Adaptation learning rate & $0.5$ \\
$\epsilon_{\mathrm{fd}}$ & Finite-difference step & $10^{-3}$ \\
Bounds & Parameter projection range & $[0.5,1.5]\times$ nominal \\
\end{tblr}

\begin{tblr}{
    row{1} = {bg=robustcolor!80,fg=black,font=\bfseries\scriptsize},
    row{3-6} = {font=\scriptsize},
    colspec = {X[0.26,c,m] X[0.5,c,m] X[0.24,c,m]},
    colsep = 1mm,
    rowsep = .2mm,
    vlines = {1pt, black},
    hlines = {1pt, black},
    hline{3-5} = {dashed,fg=gray},
}
\SetCell[c=3]{c} Robust NMPC & & \\
$m_0$ & Nominal track tightening margin & $0.10$ m \\
$\gamma$ & Mismatch-radius scaling gain & $0.02$ \\
$r_{\max}$ & Maximum mismatch radius & $0.35$ m \\
$\hat d_k$ & Additive disturbance compensation & EKF/model residual \\
\end{tblr}
\end{table}
\vspace{-0.25cm}
\bc*{\subsection{\texorpdfstring{PID Control \protect \cite{mekky2024development}}{PID Control}}
A proportional--integral--derivative (PID) controller is used as a
classical feedback baseline for raceline tracking. The steering command is computed from the tracking error and heading error, while the throttle command is computed from the velocity tracking error:
\begin{subequations}
\label{pidc}
\begin{align}
\delta = & PID(e_{\psi})+PID(e_{cte}) \\
T = & PID(e_{V_x})
\end{align}
\end{subequations}
where $PID(\cdot)$ shows the standard PID definitions with $e_{\psi}, e_{V_x}$ defined in Eq.(\ref{epsi}) and (\ref{evte}) respectively, $e_{cte}$ is the cross-track error defined as \cite{2018cte}: 
\begin{equation}
\label{cte}
    e_{cte} = (y-y_{ref})\cos{\psi_{ref}} -(x-x_{ref})\sin{\psi_{ref}}
\end{equation}
}

\subsection{\texorpdfstring{Robust Inverse of Signum Error (RISE) Control \cite{johnston2025adaptive}}{RISE Control}}
To enhance robustness against modeling uncertainty and external disturbances, a RISE-inspired nonlinear steering controller is employed. 
The controller regulates both lateral tracking accuracy and orientation alignment with respect to the reference path.
Specifically, the path-following error signal is constructed by 
combining $e_{cte}$ and $e_{V_x}$ as
\begin{equation}
r = e_{cte} + k_{e_{V_x}} e_{V_x},
\end{equation}
where $k_{e_{V_x}} > 0$ determines the relative contribution of velocity tracking error to the overall tracking objective. 
This composite error ensures that the controller simultaneously minimizes geometric deviation from the reference line while maintaining proper vehicle orientation along the path tangent. 
The steering command is then generated using a RISE-inspired feedback
structure
\begin{equation} \label{risec}
    \dot{\delta} = -k_u \norm{\delta} \mathrm{sgn}(r)-(k_s+1)r - \beta \int \mathrm{sgn}(e_{cte})
\end{equation}
where $\mathrm{sgn}(\cdot)$ represents a smoothed sign function used in
implementation. The gain $k_s$ introduces damping in steering dynamics,
$\beta$ provides instantaneous robustness to disturbances through
sign-based feedback 
and $k_u$ accumulates compensation to reject persistent uncertainties.
By explicitly coupling cross-track and velocity tracking errors in the feedback
path, the controller improves convergence during high-curvature
segments and reduces oscillatory behavior that can arise when lateral
regulation is performed independently of orientation control. 
This enables reliable path-following under sensing noise, actuation delay,
and modeling mismatch \cite{xian2004rise}.

\subsection{\texorpdfstring{Nominal Nonlinear Model Predictive Control  \cite{chrosniak2024deep}}{Nonlinear Model Predictive Control}}
Nominal Nonlinear Model Predictive Control (\nmpcc{}) is an optimization-based approach that enables racing at handling limits, but the controller explicitly uses the vehicle dynamics model to predict future system behavior over a finite horizon. At each control step, the controller solves the optimization problem
\begin{equation}
\min_{\mathbf{u}_{0:N-1}}
\sum_{k=0}^{N}
\|e_{te}{_{\vert_k}}\|_Q^2 +
\sum_{k=0}^{N-1}
\|\Delta u_k\|_R^2
\end{equation}
subject to the nonlinear vehicle dynamics as well as actuator and track constraints. 
Only the first control input is applied to the system, and the optimization is repeated at the next time step in a receding-horizon fashion. Because NMPC relies directly on the internal vehicle dynamics model, its performance is strongly influenced by model accuracy.
\vspace{-0.35cm}

\subsection{Adaptive NMPC\texorpdfstring{~\cite{hewing2020learning}}{}}
In this formulation, selected parameters of the vehicle dynamics model are updated online using a learning-based predictor trained offline from data. The updated parameters are incorporated into the prediction model at each control step, allowing the NMPC optimizer to better capture nonlinear tire behavior and actuator dynamics during aggressive maneuvering.

This approach follows the general paradigm of learning-based model predictive control, where data-driven model updates are used to enhance prediction fidelity while retaining the constraint-handling capabilities of MPC. The adaptive strategy continuously refines the model representation based on observed system evolution, enabling improved tracking performance and disturbance rejection in closed-loop racing scenarios.
Specifically, the prediction model used by NMPC is parameterized as
\begin{equation}
\label{eqn:adaptive}
x_{k+1} = f\!\left(x_k, u_k, params^*_k\right),
\end{equation}
where $params^*_k$ denotes the set of vehicle/model parameters estimated online by the learning module. These parameters are inferred from a sliding window of recent state and input measurements and are updated prior to solving the NMPC optimization problem. This learning-enhanced adaptive mechanism allows the controller to compensate for modeling errors and changing tire–road interaction conditions without introducing conservative constraint tightening.
\vspace{-0.35cm}

\subsection{Robust NMPC\texorpdfstring{~\cite{limon2010robust}}{}}

To improve closed-loop performance under modeling uncertainty and external disturbances, a disturbance-aware robust NMPC formulation is employed. The approach combines online model-mismatch estimation with constraint tightening inspired by tube-based robust MPC methods. In classical tube MPC, bounded disturbance sets are propagated to construct invariant tubes around a nominal trajectory, and optimization constraints are tightened accordingly to ensure recursive feasibility. In contrast, the present work adopts a lightweight approximation suitable for high-speed autonomous racing, where a scalar mismatch radius is computed online and used to adapt the track safety margin.
Specifically, the tightening margin is defined as
\begin{equation}
m_k = m_0 + \gamma r_k
\end{equation}
where $m_0$ denotes the nominal track safety margin, $\gamma$ is a scalar tightening gain, and $r_k$ is an online-estimated mismatch radius representing the bounded position prediction error. In practice, the mismatch radius is clipped as $r_k \in [0, r_{\max}]$ to prevent excessive constraint tightening. This adaptive tightening mechanism provides a computationally efficient surrogate for tube-based constraint tightening while preserving feasibility under moderate uncertainty.

In addition, an additive disturbance estimate is obtained from the discrepancy between the predicted model state and the EKF-estimated state,
\begin{equation}
\hat d_k =
\frac{x^{\mathrm{EKF}}_{k} - x^{\mathrm{model}}_{k}}{T_s}
\end{equation}
This disturbance term represents an estimate of unmodeled state-rate mismatch and is incorporated directly into the continuous-time prediction dynamics used by Adaptive NMPC in Eq.(\ref{eqn:adaptive}),
\begin{equation}
\hat x_{k} = f(x_k,u_k,params^*_k) + \hat d_k
\end{equation}
The discretized prediction model therefore becomes
\begin{equation}
x_{k+1}
=
x_k
+
T_s \big(f(x_k,u_k,params^*_k) + \hat d_k \big),
\end{equation}
which allows the controller to partially compensate for unmodeled dynamics and external disturbances during operation. The proposed robust NMPC formulation is inspired by tube-based MPC approaches that tighten constraints to maintain feasibility under bounded uncertainty. Unlike classical tube MPC methods that propagate invariant uncertainty sets, the present work employs an online mismatch-based tightening mechanism and an EKF-based mismatch information, suitable for real-time autonomous racing.
\vspace{-0.25cm}

\section{Benchmark Framework and Disturbance Modeling}
\label{benchmark}
\bc*{The CAPE framework is designed as a reusable controller--model benchmarking pipeline with four adjustable components:
1) Heterogeneous learned-dynamics model and controller interface; 2) Fixed train/test track split; 3) Common raceline generation; 4) Controlled uncertainty module (with $\eta$ as global noise scaling factor).   
This structure allows controller architecture, learned dynamics model, reference trajectory, and uncertainty level to be varied independently while all other benchmark settings are held fixed.}
Furthermore, to evaluate the robustness of model-based controllers, we construct a
benchmark framework that introduces realistic sensing, actuation, and
modeling uncertainties during closed-loop simulation.
\vspace{-0.25cm}

\subsection{Measurement Noise and Process Disturbances}

To evaluate controller robustness under realistic sensing, actuation,
and modeling imperfections, stochastic perturbations are introduced in
both measurement and plant dynamics during closed-loop simulation.
Multiplicative zero-mean Gaussian noise is injected into the state measurements vector
\(
\textbf{y} = [x,\; y,\; \theta,\; v_x,\; v_y,\; \omega]
\),
with standard deviation proportional to the instantaneous state
magnitude:
\begin{equation}
z_i = y_i + \mathcal{N}\!\left(0, (\sigma_i |y_i|)^2 \right),
\end{equation}
where $\sigma_i$ denotes the relative noise standard deviation coefficient associated with measurement $y_i$. For example,
$\sigma_i = 0.03$ corresponds to measurement noise with standard
deviation equal to $3\%$ of the current state magnitude. Small
regularization constants prevent the noise magnitude from vanishing
near zero.
\subsection{Process disturbances}
Unmodeled dynamics and environmental effects are represented by
additive stochastic perturbations applied after state propagation:
\begin{equation}
\textbf{x}_{k+1}^{\mathrm{dist}} = \textbf{x}_{k+1} + w_k, \qquad
w_k \sim \mathcal{N}\!\left(0, \Sigma_d(\textbf{x}_{k+1}) \right),
\end{equation}
where $\Sigma_d(\cdot)$ is a diagonal covariance whose entries scale
with the magnitudes of the dynamic states, producing disturbances in
velocity and yaw-rate dynamics.
 
\subsubsection{Fixed additional uncertainty sources}
To emulate realistic autonomous racing conditions, further uncertainty
mechanisms are incorporated:

\begin{itemize}
\item \textit{Steering delay:}
A discrete-time delay of approximately $20$\,ms is imposed
on steering commands using a FIFO buffer, capturing actuator and
communication latency.

\item \textit{IMU bias drift:}
Slowly varying sensor bias is modeled as a random walk
\(
b_{k+1} = b_k + \nu_k
\),
with $\nu_k$ zero-mean Gaussian noise, affecting heading and yaw-rate
measurements.

\item \textit{Crosswind disturbance:}
External aerodynamic perturbations are injected as
\[
V_{y_{k+1}} = V_{y_{k+1}} + \mathcal{N}(0,\sigma_{cw}^2), \quad
\omega_{k+1} = \omega_{k+1} + d_{\omega},
\]
where $\sigma_{cw} = 0.02\,\max(|v_x|,0.01)$ represents lateral
velocity disturbance with $2\%$ relative standard deviation.

\item \textit{Parametric uncertainty:}
Fixed multiplicative model mismatch is introduced by scaling vehicle
mass by $+5\%$, front tire stiffness by $-5\%$, rear tire stiffness by
$+5\%$, and effective friction limits by $-10\%$.
%
\end{itemize}

\begin{table}[H]
\centering
\caption{Uncertainty activation and scaling across global noise levels $\eta$}
\label{tab:noise_scaling}
    \begin{tblr}{
        width=\columnwidth,
          colspec={X[0.6,c,m] X[1.0,c,m] X[1.2,c,m] X[1.1,c,m] X[0.8,c,m] X[0.8,c,m] X[1.2,c,m]},
          row{1-2}={halign=c,valign=m},
          hlines={1pt, black},
          vlines,
          rowsep=1pt,
          colsep=1pt,
          stretch=0.9
        }
        $\eta(\%)$
        & {Meas. Noise(\%)} 
        & {Process Disturbance(\%)} 
        & {Actuator Delay(ms)} 
        & {IMU \\ Bias} 
        & {Cross-wind} 
        & {Parameter Uncertainty}       \\
        
        0 & 0 & 0  & 0   & off & off & off \\
        1 & 1 & 2  & 20 & on  & on  & on  \\
        2 & 2 & 4  & 20 & on  & on  & on  \\
        3 & 3 & 6  & 20 & on  & on  & on  \\
        4 & 4 & 8  & 20 & on  & on  & on  \\
        5 & 5 & 10 & 20 & on  & on  & on  \\
    \end{tblr}
\end{table}
A global noise scaling factor $\eta$ increases measurement noise intensity and process disturbance covariance across experiments, creating progressively more challenging operating conditions. Controllers are evaluated by sweeping the global noise factor $\eta \in [0,5]$ with a step size of $0.5$. 
\bc*{Each model--controller pair is evaluated over all noise levels using repeated closed-loop rollouts with different random seeds, resulting in statistically meaningful performance evaluation.
For clarity, only representative uncertainty activation levels and corresponding noise scaling at $1\%$ intervals are shown in Table~\ref{tab:noise_scaling}.}
\section{Simulation Results}
\label{sec:simulation_results}
\subsection{Training and Testing}
\bc*{To evaluate the predictive performance of the learned vehicle
dynamics models, experiments were conducted using a simulated
autonomous racing dataset generated from a 1:43 scale vehicle
simulator. The dataset~\cite{jain2021bayesrace} was sampled at 50\,Hz while a pure-pursuit controller followed a predefined raceline.}
%

\bc*{Two distinct racetracks are used in the CAPE pipeline with Track~1 (ETHZ) being used for model training, with a 90/10 train-test split, and Track~2 (ETHZMobil) being held out for the closed-loop evaluation. 
Track~2 (ETHZMobil) was kept completely unseen during training and used as the test set $\mathcal{S}_{D}^{test}$ to assess generalization and robustness across all controller–model combinations.
In this study, the ETHZMobil raceline is used only as the reference trajectory for computing closed-loop tracking errors and is not treated as ground truth.}

\subsection{Open-Loop Model Evaluation}
\bc*{Open-loop prediction accuracy of the proposed \epm{} along with \ddm{} and \dpm{} are presented in  Table.~\ref{openloop}. 
Readers are referred to see \cite{arrafilepavd} for detailed open-loop metrics and inference times of the proposed \epm{} model.}
\begin{table}[H]
\centering
\caption{Open Loop Results}
\label{openloop}
    \begin{tblr}{
        width=\columnwidth,
          colspec={X[0.6,c,m] X[1.0,c,m] X[1,c,m] X[1,c,m] },
          row{1-2}={halign=c,valign=m},
          hlines={1pt, black},
          vlines,
          rowsep=1pt,
          colsep=1pt,
          stretch=0.9,
          row{3} = {bg=EPM!60},
          row{4} = {bg=DDM!60},
          row{5} = {bg=DPM!60},
        }
        \SetCell[r=2]{c} Model & \SetCell[c=3]{} Root Mean Square Error & &  \\
         & $\boldsymbol{v_x}$ (\si{\meter\per\second}) & $\boldsymbol{v_y}$ 
        (\si{\meter\per\second}) & $\boldsymbol{\omega}$ (\si{\radian\per\second})\\
        EPM & $1.19\times10^{-4}$ & $5.43\times10^{-5}$ & $1.67\times10^{-4}$ \\
        DDM & $8.72\times10^{-5}$ & $4.02\times10^{-4}$ & $3.99\times10^{-3}$ \\
        DPM & $0.0343$ & $0.1252$ & $4.306$\\
    \end{tblr}
\end{table}
\bc*{\subsection{Fixed-Parameter Bicycle-Model comparison}}
\bc*{To isolate the effect of the learned dynamics components, we further compare the proposed learned \epm{} model against fixed-parameter bicycle-model variants using the same heterogeneous controllers. 
Specifically, the system dynamics and the internal model-based controller prediction models are replaced by fixed-parameter single-track dynamics using the corresponding \dpm{}, \ddm{}, and \epm{} coefficients. 
All models are evaluated using identical simulation setup to the learned comparisons, with the same ETHZMobil reference trajectory, reference speed, and uncertainty sweep, $\eta \in [0,5]$ with increments of $0.5$.}

\begin{figure}[H]
    \centering
    \includegraphics[width=\linewidth]{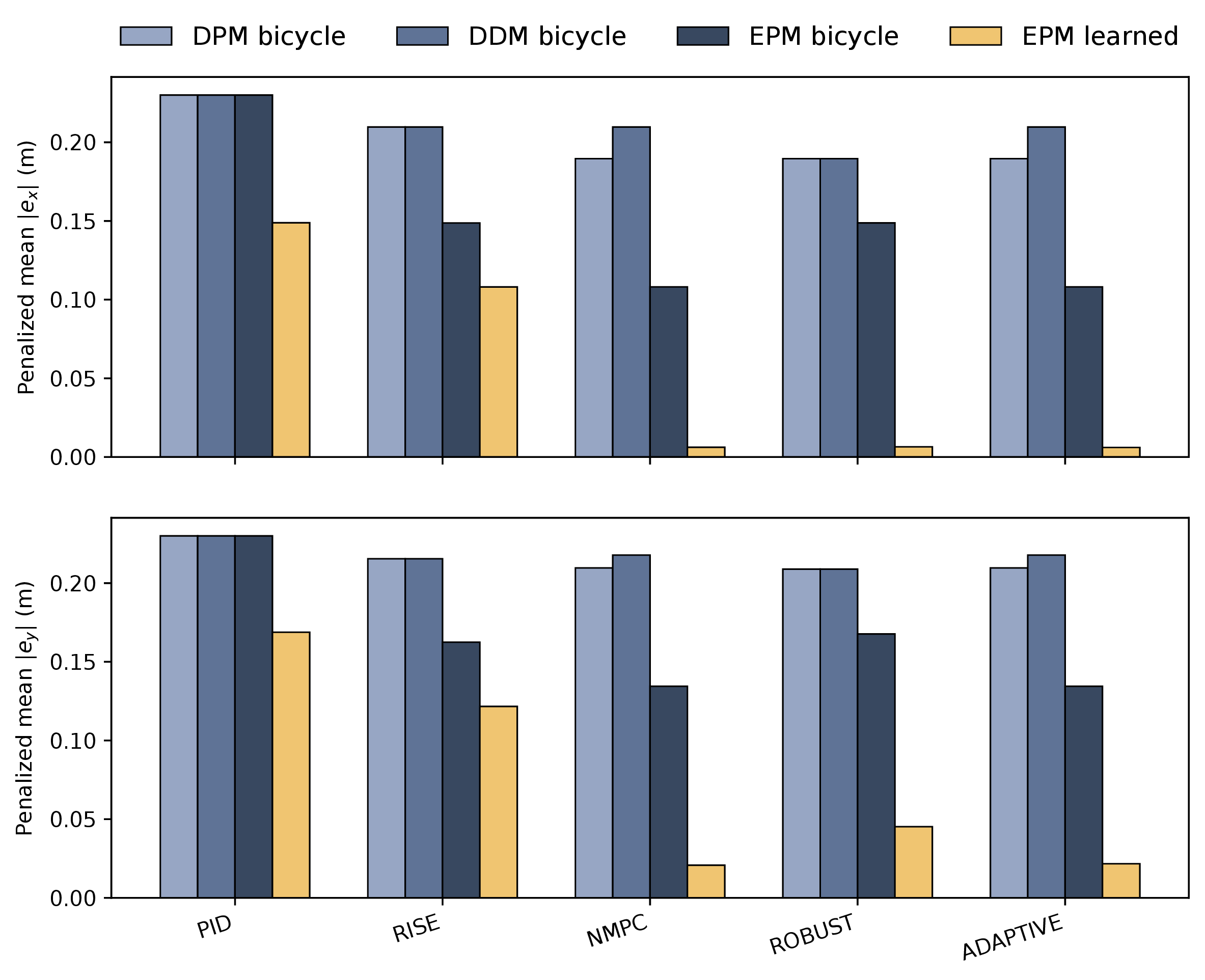}
    \vspace{-0.75\baselineskip}
    \caption{Failure-penalized tracking error comparison between fixed-parameter bicycle-model variants and learned \epm{}. Metrics are averaged over $\eta \in [0,5]$ with increments of $0.5$. 
    }
    \label{fig:bicycle_vs_epm_errors}
\end{figure}
\bc*{Fig.~\ref{fig:bicycle_vs_epm_errors} shows the failure-penalized average longitudinal and lateral tracking errors. 
Failed runs are assigned a bounded error penalty of $0.23\,\si{\meter}$, corresponding to half the ETHZMobil track width.
Across controllers, the fixed-parameter bicycle variants produce larger aggregate longitudinal and lateral errors than learned \epm{}, indicating that the single-track structure alone is insufficient for closed-loop performance under increasing levels of uncertainty.
The difference is most pronounced for the NMPC-based controllers, where learned \epm{} maintains low error across the uncertainty sweep for the Nominal, Robust, and Adaptive NMPCs.
PID fails to complete a lap for all fixed-parameter bicycle variants, while RISE fails for the added uncertainty \dpm{} and \ddm{} bicycle cases. 
\dpm{} and \ddm{} accumulate the same error due to the failure to complete a lap and further share the same fixed single-track model parameters.}
\begin{figure}[H]
    \centering
    \includegraphics[width=\linewidth]{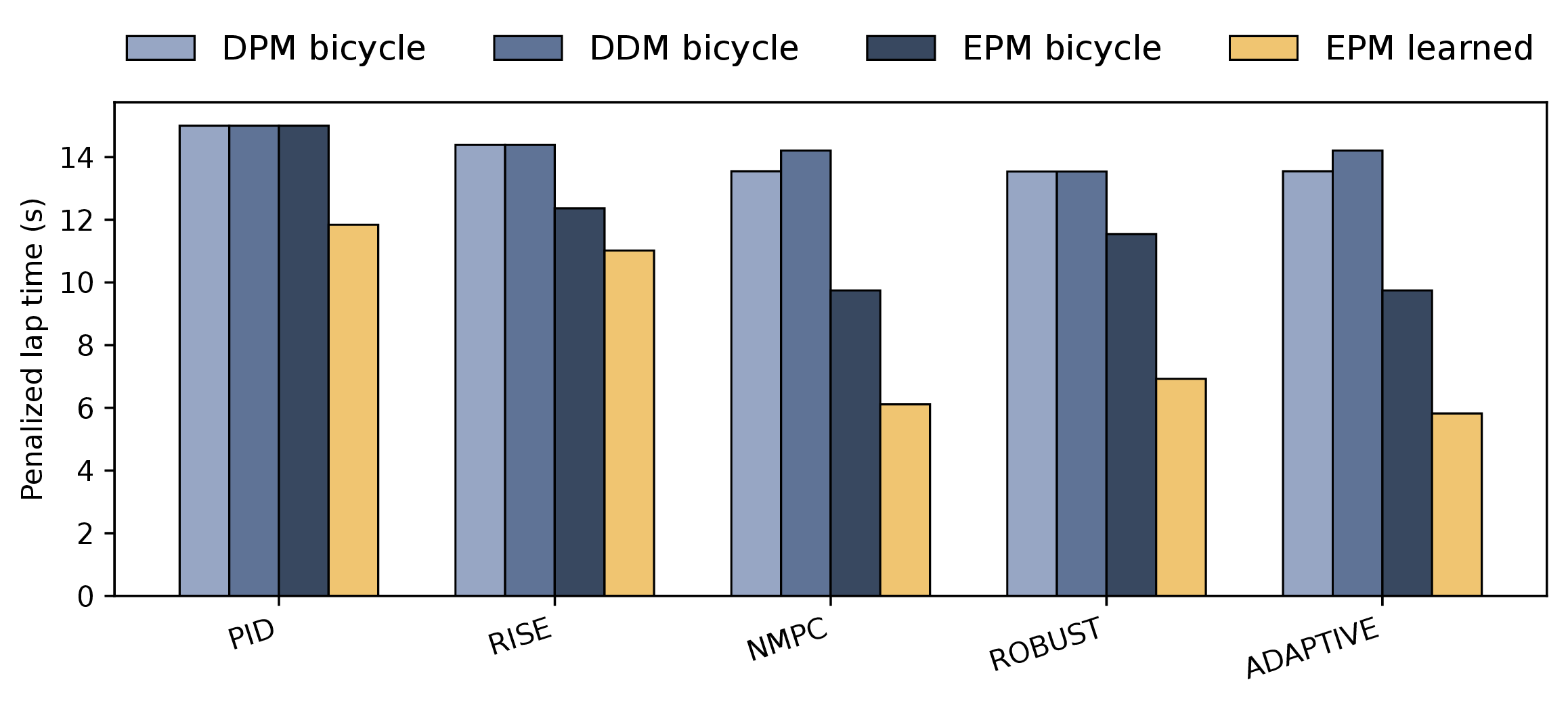}
    \vspace{-0.75\baselineskip}
    \caption{Failure-penalized lap-time comparison between fixed-parameter bicycle-model variants and learned \epm{}.}
    \label{fig:bicycle_vs_epm_laptime}
\end{figure}

\bc*{Fig.~\ref{fig:bicycle_vs_epm_laptime} compares the corresponding penalized lap-completion times. 
The fixed-parameter bicycle variants produce longer penalized lap times because many runs fail to complete a single lap.
Each failed run is assigned a bounded penalty time of $15\,\si{\second}$. 
On the other hand, learned \epm{} model completes all evaluated uncertainty levels for \nmpcc{}, \rnmpc{}, and \anmpc{}, resulting in substantially lower penalized lap times.
The results show that the learned tire, drivetrain, and transient dynamics components provide the additional model fidelity needed for robust high-speed tracking under sensing noise, process disturbances, crosswind perturbations, and parametric uncertainty.}
\bc*{Table~\ref{tab:bicycle_best_controller} shows the maximum $\eta$ tolerance for each bicycle model-controller pair and the corresponding single lap completion times.} 
\bc*{All 3 bicycle models achieve best performance with nonlinear model predictive control variants. 
The \epm{} bicycle variant achieves a lower one lap time for a moderate $\eta$ level, indicating that its higher-fidelity dynamic equations provide improved closed-loop behavior even without learned online parameter updates.}
\begin{table}[h]
    \centering
    \caption{Best Controller for each fixed-bicycle model before failure} 
    \label{tab:bicycle_best_controller}
    \begin{tblr}{
      width = \linewidth,
      colspec = {X[0.9,c] X[0.9,c] X[0.6,c] X[0.8,c]} ,
      row{1,5} = {font=\bfseries},      
      row{4,8} = {bg=EPM!60},
      row{3,7} = {bg=DDM!60},
      row{2,6} = {bg=DPM!60},
      hlines = {1pt, black},
      hline{3,4,6,7,9,10,12,13,15,16} = {fg=gray},
      vlines = {0.6pt, black},
      rowsep = 1pt,
      colsep = 1pt,
    }

Bicycle Model & Controller  & Max. $\eta$  &  One Lap - Time \\ 
\dpm{} & \nmpcc{} & 3.0 & $7.74\,\si{\second}$ \\
\ddm{} & \rnmpc{} & 0.5 & $7.32\,\si{\second}$ \\
\epm{} & \nmpcc{} & 2.5 & $5.66\,\si{\second}$ \\  

    \end{tblr}
\end{table}  

\subsection{Learned-Model Closed-Loop Results}
\begin{figure}[H]
    \centering
    \includegraphics[width=\linewidth]{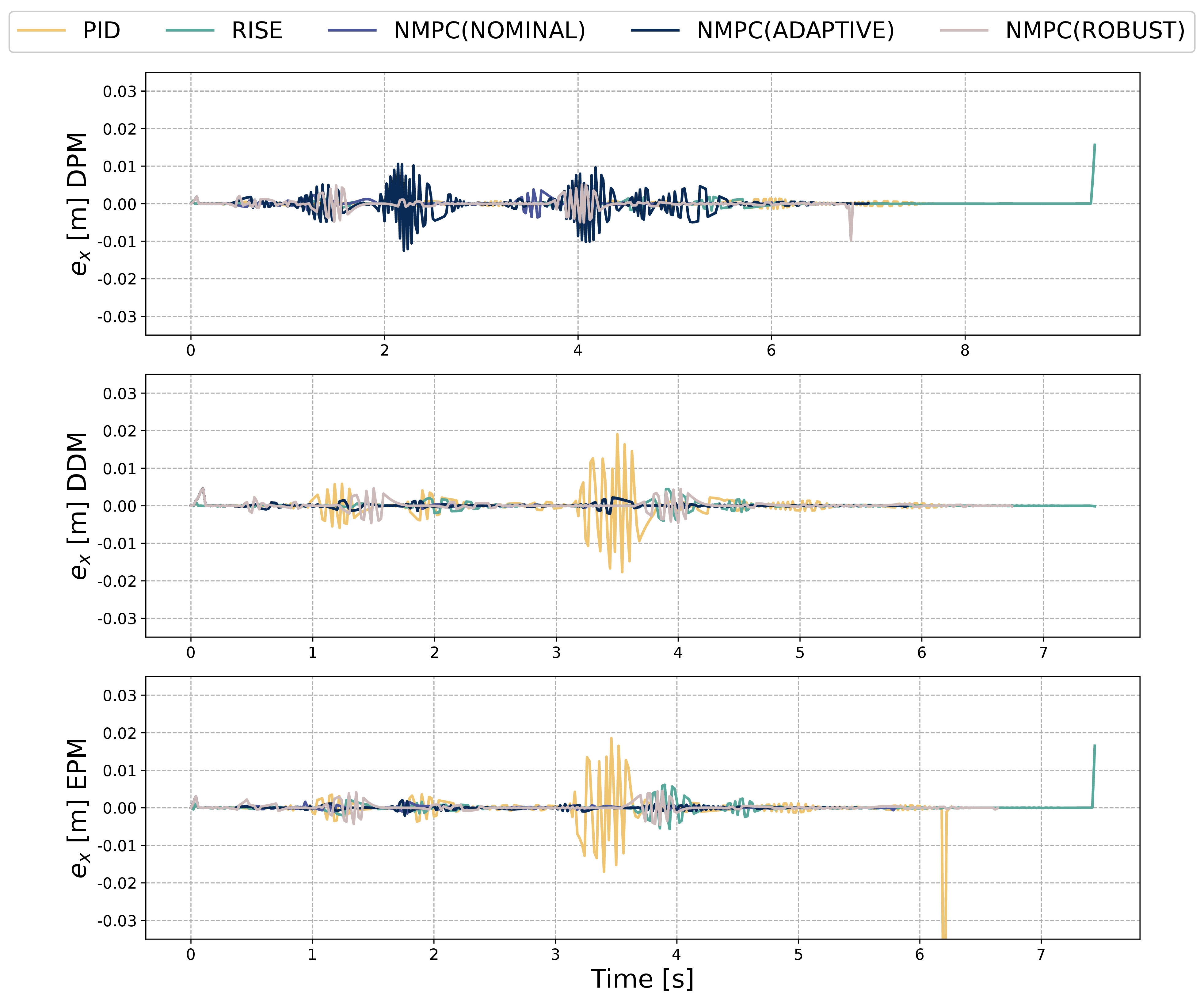}
    \vspace{-\baselineskip}
        \caption{Temporal evolution of longitudinal tracking error $e_x$.}
    \label{fig:ex}
\end{figure}

\begin{figure}[H]
    \centering
    \includegraphics[width=\linewidth]{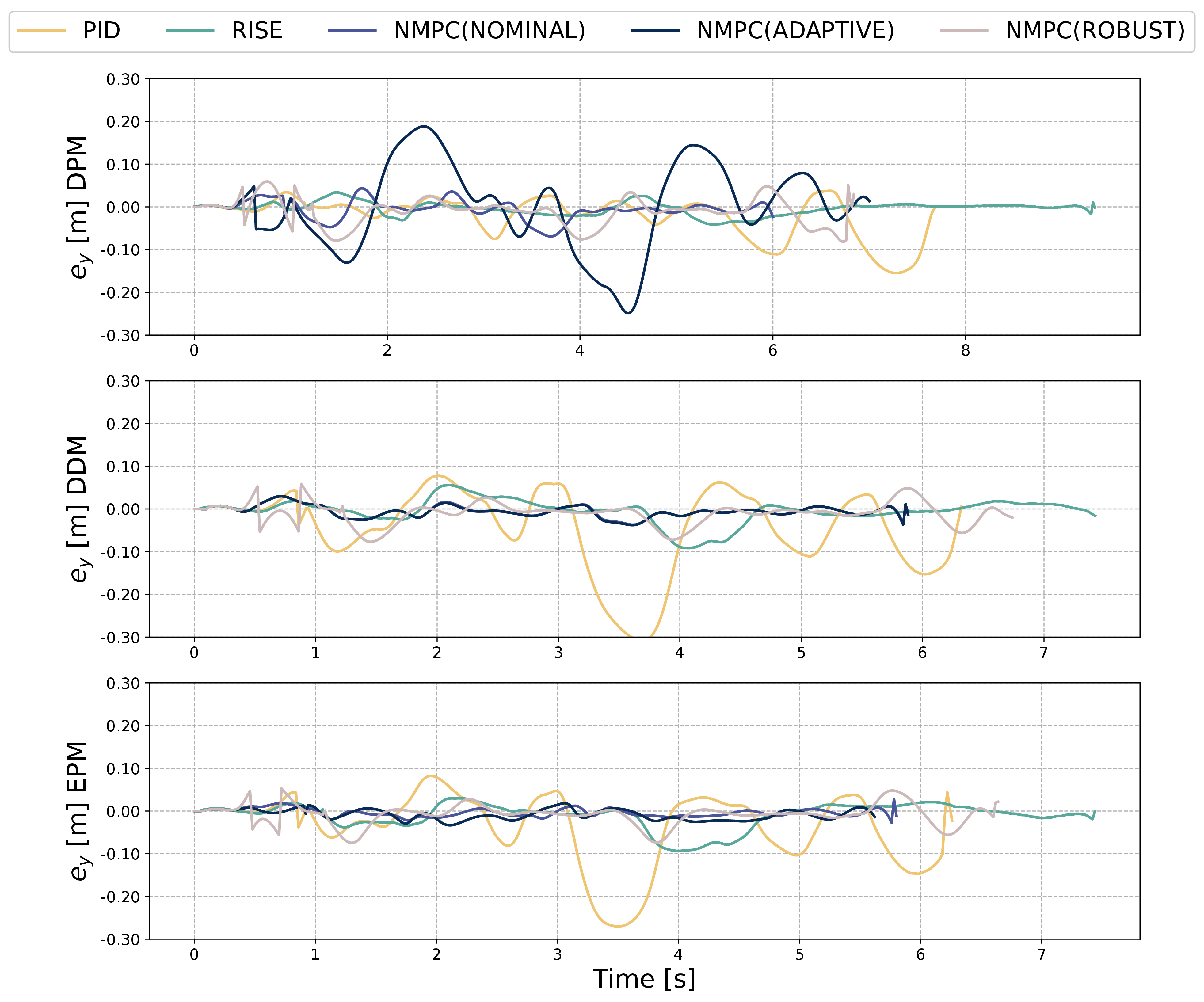}
    \vspace{-\baselineskip}
       \caption{Temporal evolution of lateral tracking error $e_y$.}
    \label{fig:ey}
\end{figure}

\bc*{Closed-loop state evolution is generated by the selected learned dynamics model with \pidc{}, \risec{}, \nmpcc{}, \rnmpc{}, \anmpc{} control architectures. 
Especially for the model-based controllers, the internal model used by the optimizer is the respective learned model.
Thus, the reported errors quantify relative model-controller behavior under identical reference, track, and uncertainty conditions.}
Fig.~\ref{fig:ex} and Fig.~\ref{fig:ey} compare the temporal evolution of longitudinal and lateral tracking errors across all 3 learned dynamics models and 5 controller architectures. 
Distinct transient behaviors are observed in terms of peak deviation, oscillation, and settling characteristics.
NMPC-based controllers exhibit more aggressive corrections during curvature transitions, while PID shows smoother but slower convergence, the RISE controller on the other hand provides consistent error attenuation with reduced sustained oscillations.  

%

\begin{figure*}[!ht]
    \centering
    \includegraphics[width=\linewidth, height=0.28\linewidth]{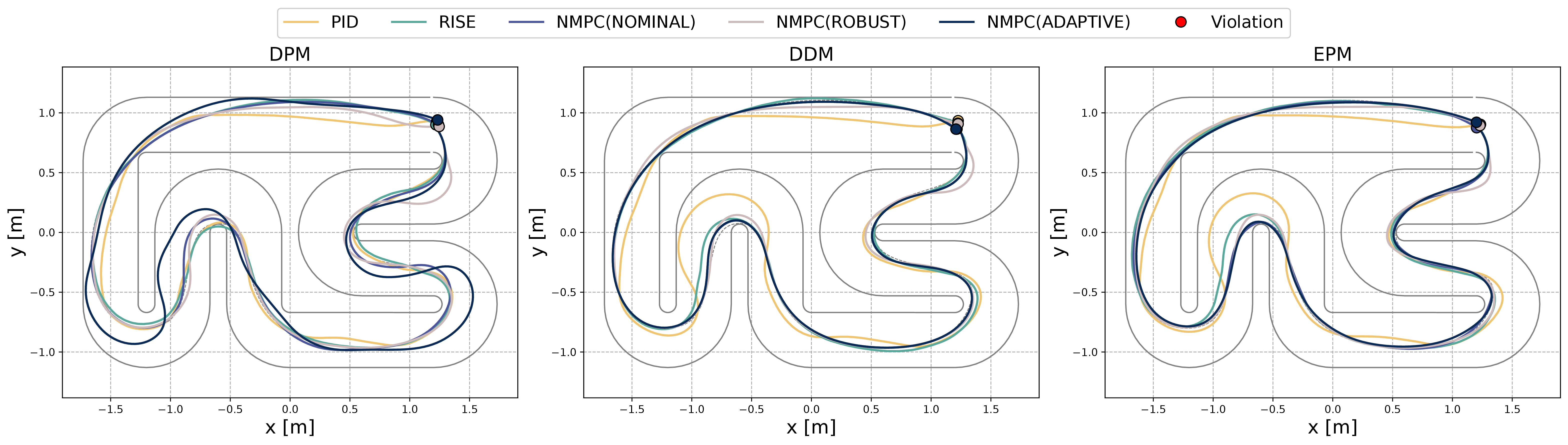}
    \vspace{-0.5\baselineskip}
    \caption{Comparison of closed-loop path-following behavior under different controllers. The left, middle, and right rows present trajectories obtained from \dpm{}- left, \ddm{}-middle, and proposed \epm{}-right models. [for $\eta=0$ case] }
    \label{fig:traj}
\end{figure*}

Fig.~\ref{fig:traj} compares the closed-loop trajectories generated by all five controllers on all 3 learned dynamics models. 
Variations in path deviation are mainly observed during high-curvature segments, where controller aggressiveness and model fidelity influence race-line adherence.
NMPC-based strategies follow the reference more tightly but exhibit sharper corrective maneuvers, whereas PID produces smoother yet less precise tracking and RISE maintains consistent trajectory regulation with moderate correction effort.



\begin{figure}[t]
    \centering
    \includegraphics[width=\linewidth, height=\linewidth]{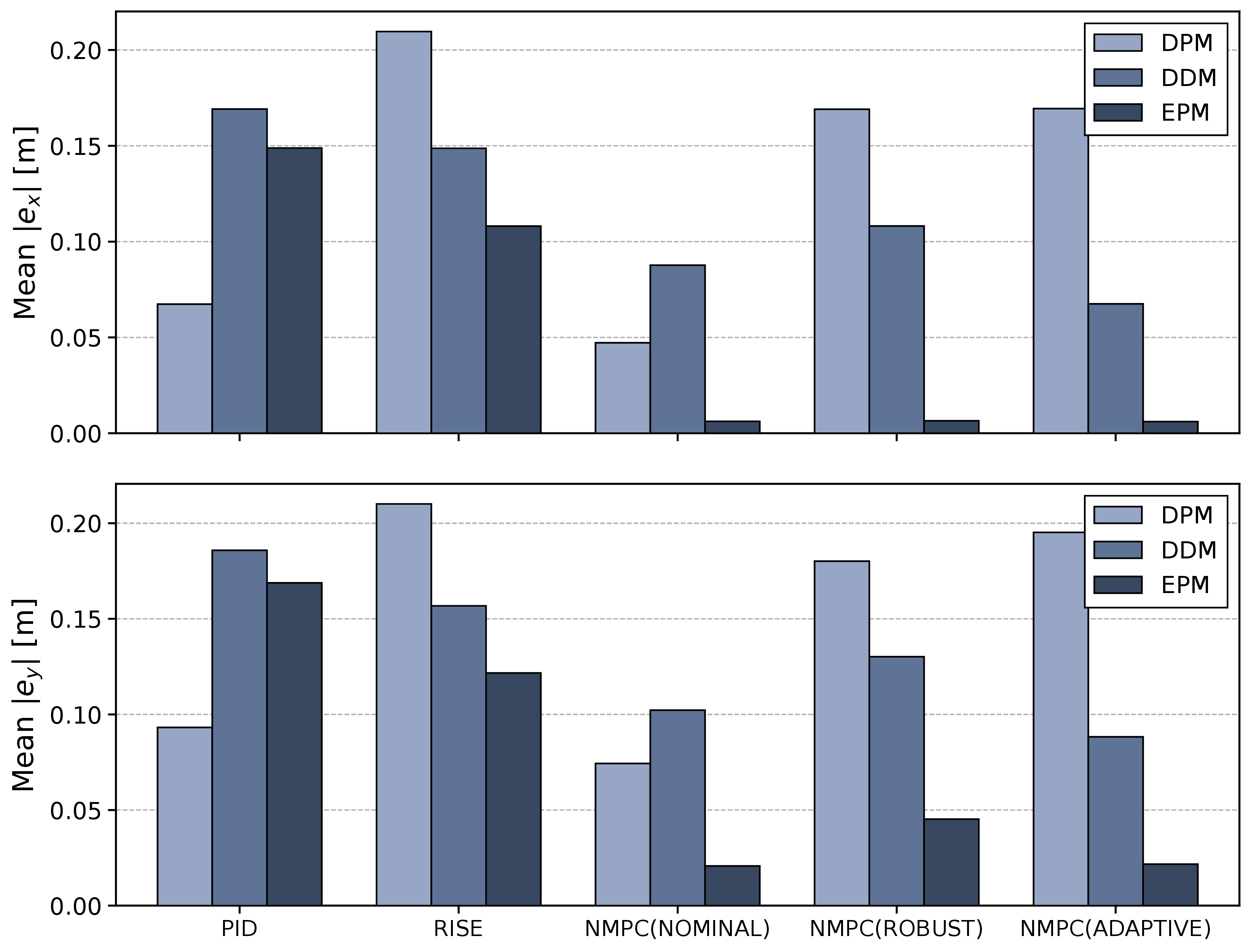}
    \vspace{-0.75\baselineskip}
    \caption{Failure-penalized average longitudinal tracking error $\mathrm{Mean}\,|e_x|$ (top) and lateral tracking error $\mathrm{Mean}\,|e_y|$ (bottom), averaged over $\eta \in [0,5]$ with increments of $0.5$.} 
    \label{fig:mean_errors_all_ctrl}
\end{figure}
\bc*{Fig.~\ref{fig:mean_errors_all_ctrl} shows the failure-penalized average tracking errors $(e_x,e_y)$ for all 5 controllers and 3 learned models across the full uncertainty sweep averaged over $\eta \in [0,5]$ with increments of $0.5$. Failed runs are assigned a bounded tracking-error penalty of half the track width equal to $0.23\,\mathrm{m}$.
%
Across the feedback-based \pidc{} and \risec{} controllers, the differences between all three learned models are less pronounced since these controllers do not explicitly optimize over a prediction model.
For \pidc{}, the lower \dpm{} tracking error is due to the conservative steering/reference-speed regulation without throttle optimization, which reduces deviation but leads to longer lap times, whereas with \risec{}, \epm{} achieves the lowest aggregate longitudinal and lateral tracking errors. 
The advantage of \epm{} becomes more evident for the NMPC-based controllers, where prediction accuracy directly affects the planned trajectory and control inputs.
For nominal, robust, and adaptive NMPC, \epm{} consistently produces the lowest aggregate longitudinal and lateral tracking errors, indicating improved model fidelity and greater closed-loop robustness under increasing uncertainty.
The larger penalized errors observed for the \dpm{} and \ddm{} variants are driven both by increased tracking deviation during completed laps and by failure penalties incurred at higher noise levels.}
\begin{figure}[t]
    \centering
    \includegraphics[width=\linewidth, height=\linewidth]{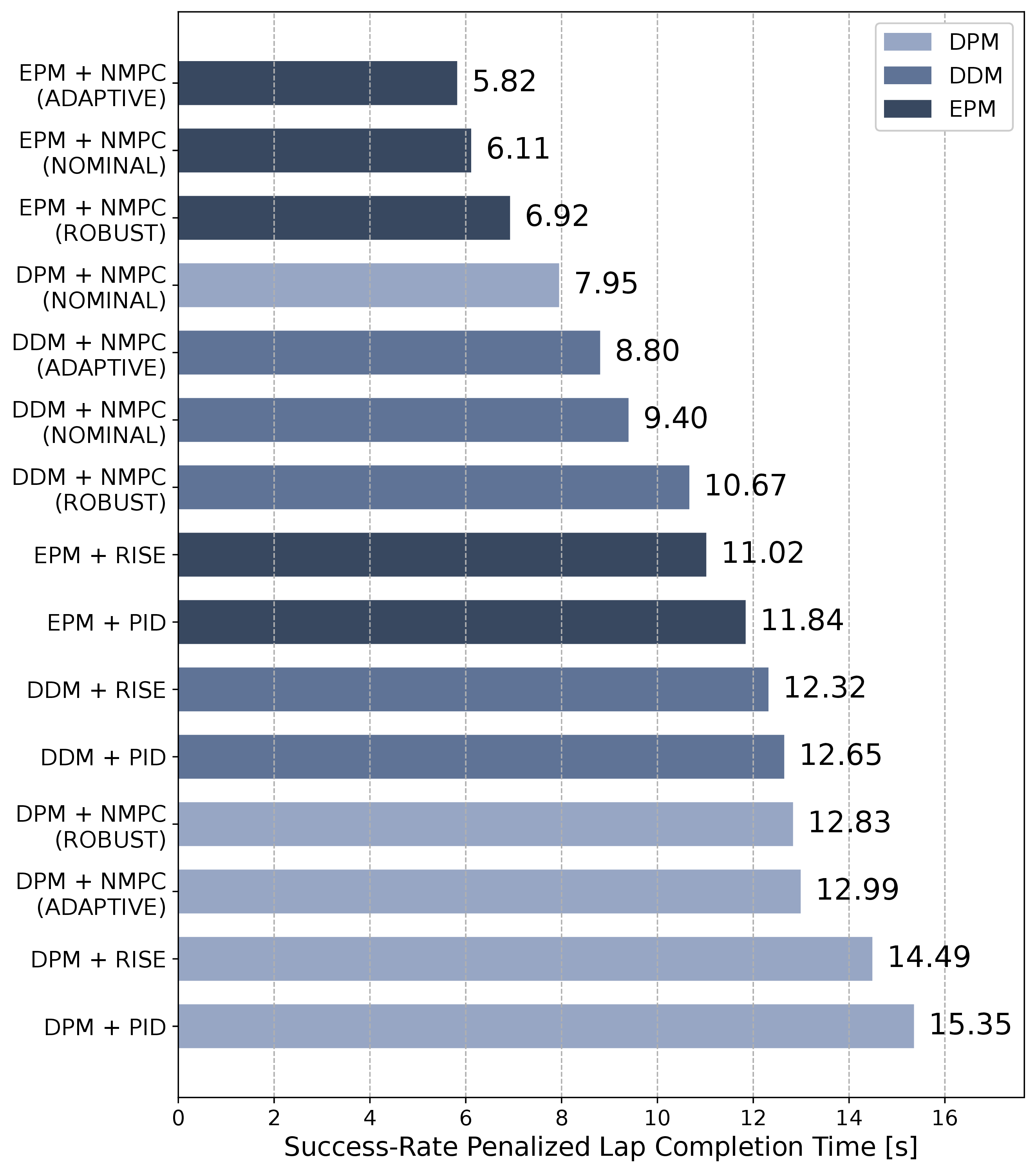}
    \vspace{-0.5\baselineskip}
    \caption{Comparison of lap times for controller–model performance using overall penalized lap completion time. 
    }
    \label{fig:finishtime}
\end{figure}

Fig.~\ref{fig:finishtime} summarizes the overall penalized lap completion times for all 15 model-controller combinations.
The reported metric assigns duration of $15\,\mathrm{sec}$ to unsuccessful runs.
Variations in completion time reflect differences in tracking accuracy, control smoothness, and motion consistency observed in earlier analyses. 
\bc*{Overall NMPC-based controllers paired with the proposed \epm{} achieve consistent faster lap completion times with fewer track violations indicating improved closed-loop performance as $\eta$ increases}. 
\bc*{For example, with \anmpc{}, \epm{} completes the lap in $5.82\,\text{s}$, compared with $12.99\,\text{s}$ for \dpm{} and $8.80\,\text{s}$ for \ddm{}.}
In contrast, PID and RISE exhibit longer execution times due to more conservative regulation, reduced ability to maintain high-speed tracking, \bc*{and finally failing to even complete the lap at all as $\eta$ increases}.

\begin{table}[H]
    \centering
    \caption{Closed-loop performance comparison under moderate sensing uncertainty ($\eta = 1$)} 
    \label{tab:noise}
    \begin{tblr}{
      width = \linewidth,
      colspec = {X[0.7,c] X[0.7,c] *{4}{X[0.7,c]}} ,
      row{1} = {font=\bfseries},      
      row{2-4} = {bg=pidcolor!40},
      row{5-7} = {bg=risecolor!40},
      row{8-10} = {bg=nmpccolor!40},
      row{14-16} = {bg=robustcolor!40},
      row{11-13} = {bg=adaptivecolor!40},
      hlines = {1pt, black},
      hline{3,4,6,7,9,10,12,13,15,16} = {fg=gray},
      vlines = {0.6pt, black},
      rowsep = 1pt,
      colsep = 1pt,
    }

Controller & Dynamics Model & $\boldsymbol{e_x}$ (\si{\meter}) 
& $\boldsymbol{e_y}$ (\si{\meter})   
& $\sqrt{V_x^2+V_y^2}$ (\si{\meter\per\second})
& Lap time (\si{\second}) \\

\SetCell[r=3]{c} PID 
& \dpm{}   & 0.0194 & 0.0317 & 0.660 & 16.260 \\
& \ddm{}   & 0.0403 & 0.0402 & 1.847 & 6.280 \\
& \epm{}   & 0.0412 & 0.0457 & 1.840 & 6.440 \\

\SetCell[r=3]{c} RISE  
& \dpm{}   & 0.0082 & 0.0052 & 0.833 & Failed \\
& \ddm{}    & 0.0160 & 0.0201 & 1.468 & 7.660 \\ 
& \epm{} & 0.0179 & 0.0240 & 1.462 & 7.720 \\

\SetCell[r=3]{c} Nominal NMPC  
& \dpm{}   & 0.0164 & 0.0183 & 1.822 & 6.140 \\
& \ddm{}    & 0.0197 & 0.0239 & 1.674 & Failed \\ 
& \epm{} & 0.0079 & 0.0104 & 1.834 & 5.920 \\

\SetCell[r=3]{c} Adaptive NMPC  
& \dpm{}   & 0.0666 & 0.0637 & 1.261 & Failed \\
& \ddm{}     & 0.0172 & 0.0192 & 1.829 & 6.100 \\
& \epm{} & 0.0096 & 0.0097 & 1.901 & 5.640 \\

\SetCell[r=3]{c} Robust NMPC  
& \dpm{}   & 0.0381 & 0.0414 & 1.661 & 7.220 \\
& \ddm{}     & 0.0343 & 0.0343 & 1.646 & 7.080 \\
& \epm{} & 0.0290 & 0.0310 & 1.688 & 6.800 \\

    \end{tblr}
\end{table}  
\begin{figure*}[b]
    \centering
    \includegraphics[width=\linewidth,height=0.763\linewidth,keepaspectratio]{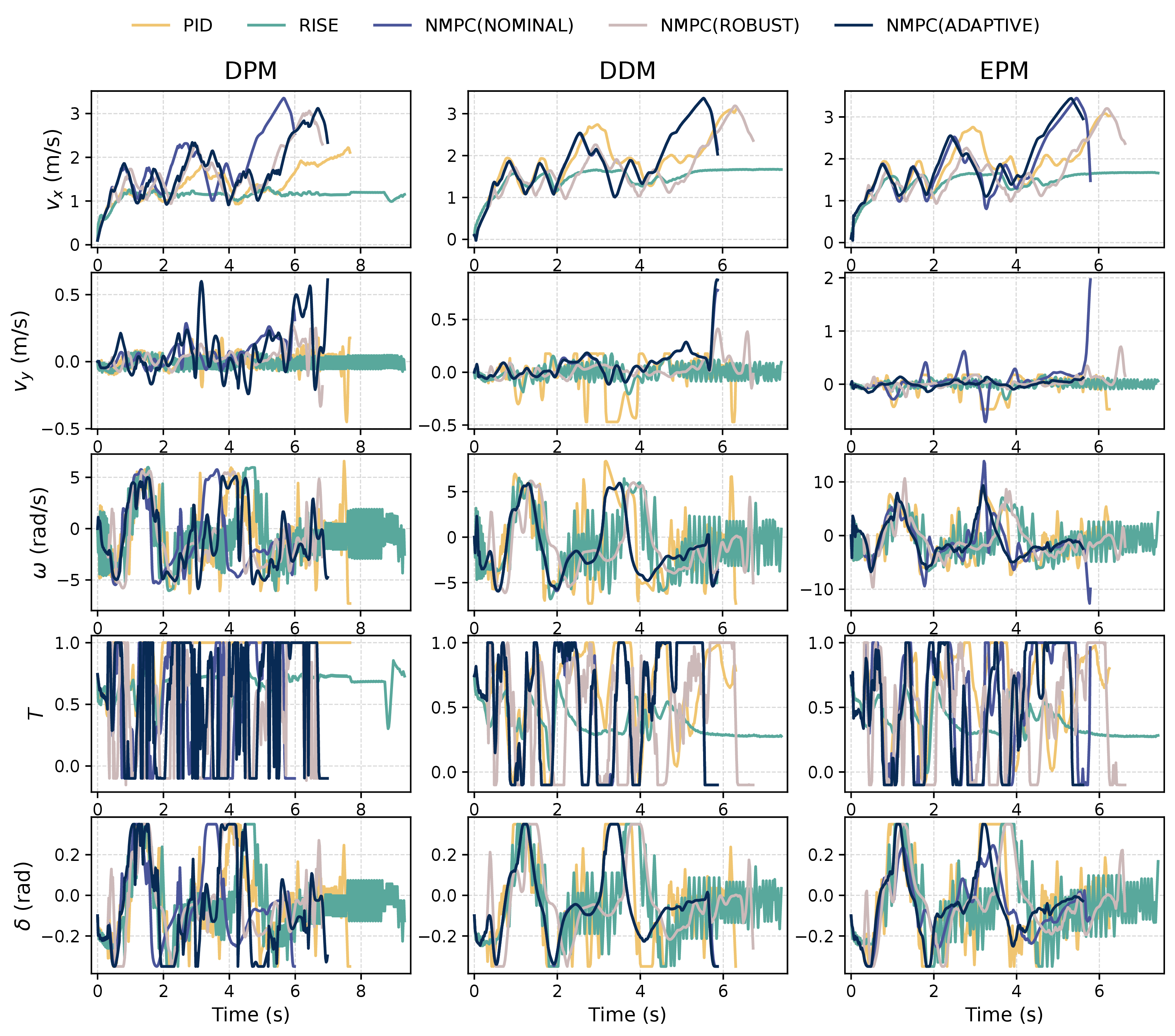}
    \vspace{-0.5\baselineskip}
    \caption{Each row shows the closed-loop longitudinal velocity $v_x$, lateral velocity $v_y$, yaw rate $\omega$, throttle command $T$, and steering command $\delta$ respectively for all 5 different controllers evaluated on \dpm{} (left), \ddm{} (middle), and the proposed \epm{} (right) respectively under $\eta = 0$ case.}
    \label{fig:vx_vy_yaw}
\end{figure*}

Table~\ref{tab:noise} reports the closed-loop performance metrics for a global noise factor value $\eta=1$. 
For each controller, the table provides the mean longitudinal and lateral tracking errors ($e_x$, $e_y$), mean velocity magnitude  $\sqrt{V_x^2+V_y^2}$, and lap completion time, for all 3 learned models and 5 controllers.
Controllers using the \epm{} consistently demonstrate improved robustness, with lower tracking errors and successful lap completions in scenarios where \dpm{}, \ddm{} based controllers fail.
As shown in Table~\ref{tab:model_complexity}, the reduced complexity of \dpm{} correlates with reduced robustness under uncertainty.
While \dpm{} produces results in clean conditions, the introduction of measurement noise rapidly degrades tracking performance.
These results indicate that improvements in the internal vehicle dynamics representation can significantly enhance the robustness and performance of closed-loop autonomous racing controllers under a myriad of uncertain conditions.
%
\bc*{Averaged across the five controllers at $\eta=1$, \epm{} reduces longitudinal tracking error by $29.0\%$ and $17.2\%$, lateral tracking error by $24.6\%$ and $12.3\%$, and increases average velocity magnitude by $39.9\%$ and $3.1\%$, relative to \dpm{} and \ddm{}, respectively. 
For feedback-only \pidc{} and \risec{} controllers, \ddm{} cases exhibit marginally higher average speed than \epm{} because these controllers do not optimize over the learned prediction model, so the small speed difference mainly reflects closed-loop trajectory variation rather than improved predictive robustness.}
%

%

\begin{figure*}[!ht]
    \centering
    \includegraphics[width=\linewidth, height=0.9\linewidth,keepaspectratio]{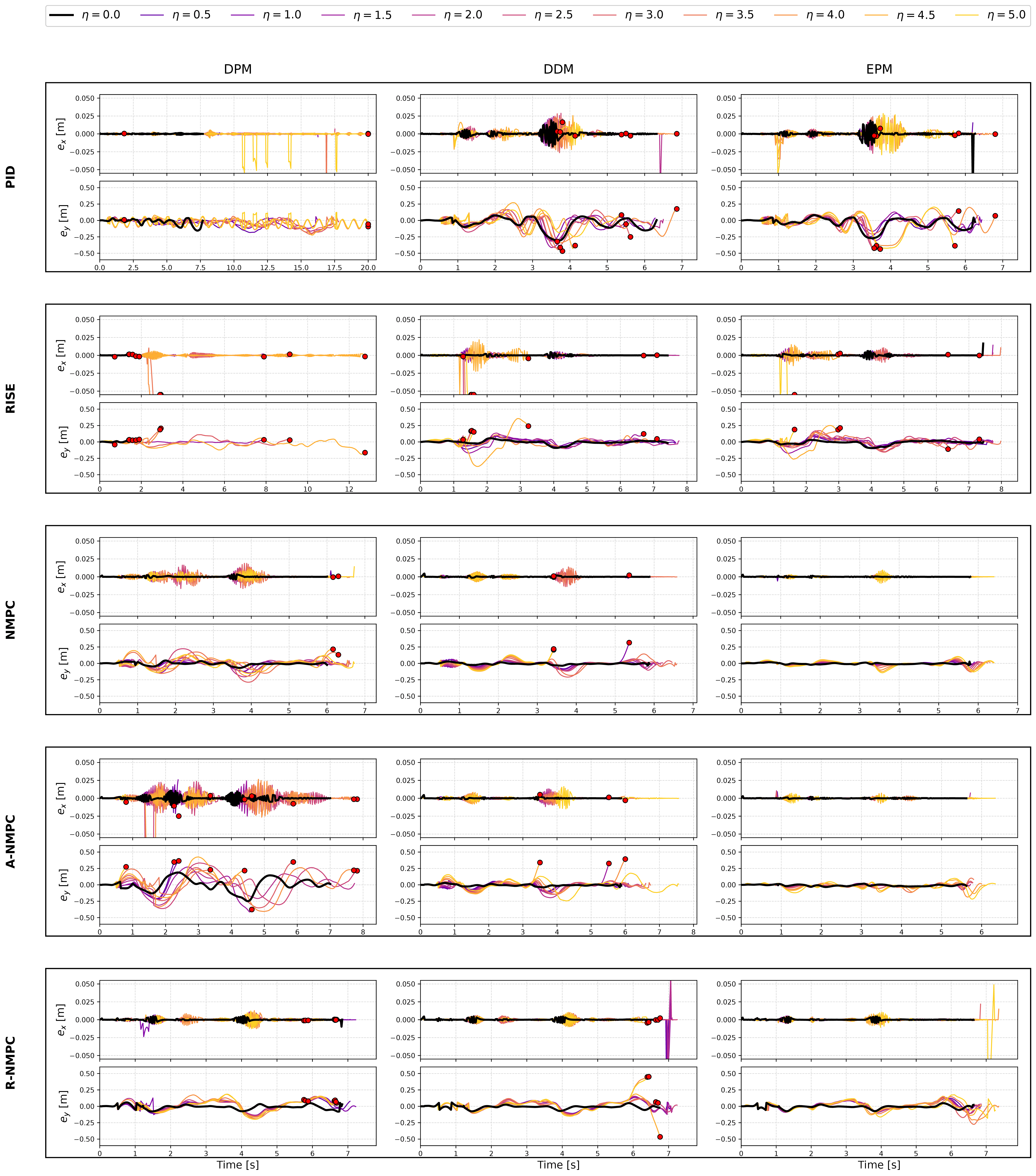}
    \vspace{-0.5\baselineskip}
    \caption{Temporal evolution of longitudinal and lateral tracking errors of the five controllers over the uncertainty sweep $\eta \in [0,5]$ with $0.5$ increments.}
    \label{fig:eta_sweep}
\end{figure*}
\bc*{Fig.\ref{fig:vx_vy_yaw} compares the closed-loop dynamic-state
responses for all 15 controller--model combinations under clean conditions.
Across all models, the NMPC-based controllers reach higher longitudinal
velocities than PID and RISE, while producing larger lateral-velocity and
yaw-rate variations during high-curvature portions of the lap.
\ddm{} with \nmpcc{} and \anmpc{} controllers provide only marginally improvements relative to its nominal behavior, indicating that its fixed-load Pacejka-style parameterization
provides limited state-dependent adaptation in the clean rollout.
In contrast, the proposed \epm{} exhibits stronger speed and curvature dependent responses due to the incorporation of load-sensitive tire forces, dynamic load-transfer and actuator-aware parameterization changes with throttle, steering, along the raceline.
This leads to larger yaw-rate responses during aggressive cornering, especially for the NMPC-based controllers, while still preserving stable tracking and faster lap completion.}
PID yields smoother but less adaptive responses, while RISE maintains moderated control effort with reduced sustained oscillations. Notably, the proposed \epm{} produces more temporally consistent control inputs, smoother vehicle motion progression and achieves fast lap times across all controllers, \bc*{indicating improved closed-loop performance with reduced sensitivity to aggressive control actions}. 

\bc*{Fig.~\ref{fig:eta_sweep} compares the temporal evolution of the longitudinal and lateral tracking errors for three learned dynamics models and five controller architectures over $\eta \in [0,5]$ with $0.5$ increments, evaluating how each controller-model pair responds as the CAPE disturbance pipeline progressively increases sensing, actuation, process, and parametric uncertainties. As $\eta$ increases, the errors show larger transient deviations and oscillatory behavior. However, \epm{} maintains lower longitudinal and lateral error magnitudes than \dpm{} and \ddm{} across most controllers, indicating that its higher-fidelity dynamics representation improves closed-loop state propagation and NMPC prediction.}
\begin{table*}[!ht]
\centering
\caption{Controller success and lap completion time across various global noise factor ($\eta$) levels for all 3 learned models (\dpm{}, \ddm{} \& \epm{})}
\label{tab:noise_ablation}

\newcommand{\cmark}{\ding{51}} 
\newcommand{\xmark}{\footnotesize{\ding{55}}} 
\begin{tblr}{
    width=\linewidth,
    colspec={
        X[0.8,c,m]
        X[0.6,c,m]
        *{11}{X[0.22,c,m] X[0.35,c,m]}
    },
    row{1-3}={font=\bfseries},
    row{4-6}={bg=pidcolor!40},
    row{7-9}={bg=risecolor!40},
    row{10-12}={bg=nmpccolor!40},
    row{13-15}={bg=adaptivecolor!40},
    row{16-18}={bg=robustcolor!40},
    hline{1,Z}={0.9pt},
    hline{4,7,10,13,16}={0.7pt},
    hline{2,3}={0.4pt},
    vline{1,2,3,Z}={0.7pt},
    vline{5,7,9,11,13,15,17,19,21,23}={0.35pt},
    rowsep=1pt,
    colsep=1.5pt,
}
\SetCell[r=3]{c} Controller & \SetCell[r=3]{c} Model & \SetCell[c=22]{c} $\eta$ \\

&  & \SetCell[c=2]{c} 0 & &\SetCell[c=2]{c} 0.5
& & \SetCell[c=2]{c} 1
& & \SetCell[c=2]{c} 1.5
& &\SetCell[c=2]{c} 2
& & \SetCell[c=2]{c} 2.5
& & \SetCell[c=2]{c} 3
& & \SetCell[c=2]{c} 3.5
& & \SetCell[c=2]{c} 4
& & \SetCell[c=2]{c} 4.5
& & \SetCell[c=2]{c} 5
\\

&
&
S & T
& S & T
& S & T
& S & T
& S & T
& S & T
& S & T
& S & T
& S & T
& S & T
& S & T
\\

\SetCell[r=3]{c} PID
& DPM
& \cmark & 7.68 & \cmark & 12.64 & \cmark & 16.26
& \xmark & -- & \cmark & 17.40 & \cmark & 17.52
& \cmark & 17.80 & \cmark & 17.10 & \xmark & --
& \cmark & 17.50 & \xmark & -- \\

& DDM
& \cmark & 6.32 & \xmark & -- & \cmark & 6.28
& \xmark & -- & \cmark & 6.50 & \xmark & --
& \xmark & -- & \xmark & -- & \xmark & --
& \xmark & -- & \xmark & -- \\

& EPM
& \cmark & 6.26 & \cmark & 6.20 & \cmark & 6.44
& \cmark & 6.34 & \xmark & -- & \xmark & --
& \xmark & -- & \xmark & -- & \xmark & --
& \xmark & -- & \xmark & -- \\

\SetCell[r=3]{c} RISE
& DPM
& \cmark & 9.34 & \xmark & -- & \xmark & --
& \xmark & -- & \xmark & -- & \xmark & --
& \xmark & -- & \xmark & -- & \xmark & --
& \xmark & -- & \xmark & -- \\

& DDM
& \cmark & 7.42 & \cmark & 7.64 & \cmark & 7.66
& \xmark & -- & \cmark & 7.76 & \xmark & --
& \xmark & -- & \xmark & -- & \xmark & --
& \xmark & -- & \xmark & -- \\

& EPM
& \cmark & 7.44 & \cmark & 7.60 & \cmark & 7.72
& \cmark & 7.74 & \cmark & 7.74 & \xmark & --
& \xmark & -- & \cmark & 7.98 & \xmark & --
& \xmark & -- & \xmark & -- \\

\SetCell[r=3]{c} {Nominal\\NMPC}
& DPM
& \cmark & 6.00 & \cmark & 6.12 & \cmark & 6.14
& \cmark & 6.38 & \cmark & 6.32 & \cmark & 6.54
& \cmark & 6.62 & \xmark & -- & \cmark & 6.58
& \xmark & -- & \cmark & 6.72 \\

& DDM
& \cmark & 5.88 & \cmark & 5.98 & \xmark & --
& \cmark & 6.08 & \cmark & 6.16 & \cmark & 6.22
& \cmark & 6.46 & \cmark & 6.58 & \xmark & --
& \xmark & -- & \xmark & -- \\

& EPM
& \cmark & 5.80 & \cmark & 5.90 & \cmark & 5.92
& \cmark & 6.00 & \cmark & 6.04 & \cmark & 6.10
& \cmark & 6.16 & \cmark & 6.24 & \cmark & 6.30
& \cmark & 6.38 & \cmark & 6.40 \\

\SetCell[r=3]{c} {Adaptive\\NMPC}
& DPM
& \cmark & 7.00 & \xmark & -- & \xmark & --
& \xmark & -- & \cmark & 7.74 & \xmark & --
& \cmark & 8.14 & \xmark & -- & \xmark & --
& \xmark & -- & \xmark & -- \\

& DDM
& \cmark & 5.88 & \cmark & 6.06 & \cmark & 6.10
& \xmark & -- & \cmark & 6.40 & \cmark & 6.50
& \cmark & 6.60 & \cmark & 6.74 & \xmark & --
& \xmark & -- & \cmark & 7.56 \\

& EPM
& \cmark & 5.62 & \cmark & 5.56 & \cmark & 5.64
& \cmark & 5.68 & \cmark & 5.72 & \cmark & 5.76
& \cmark & 5.78 & \cmark & 5.80 & \cmark & 6.00
& \cmark & 6.12 & \cmark & 6.34 \\

\SetCell[r=3]{c} {Robust\\NMPC}
& DPM
& \cmark & 6.84 & \cmark & 7.06 & \cmark & 7.22
& \xmark & -- & \xmark & -- & \xmark & --
& \xmark & -- & \xmark & -- & \xmark & --
& \xmark & -- & \xmark & -- \\

& DDM
& \cmark & 6.74 & \cmark & 7.04 & \cmark & 7.08
& \cmark & 7.12 & \cmark & 7.10 & \cmark & 7.24
& \xmark & -- & \xmark & -- & \xmark & --
& \xmark & -- & \xmark & -- \\

& EPM
& \cmark & 6.64 & \cmark & 6.86 & \cmark & 6.80
& \cmark & 6.76 & \cmark & 6.82 & \cmark & 6.86
& \cmark & 6.84 & \cmark & 7.10 & \cmark & 7.36
& \cmark & 6.86 & \cmark & 7.26 \\

\end{tblr}
\end{table*}
\bc*{Failure markers (red dots) further show that learned \epm{} completes all NMPC runs, with $0$ failures across \nmpcc{}, \rnmpc{} and \anmpc{}.
\dpm{} and \ddm{} accumulate $19$ and $12$ failures,
respectively, across the same $33$ NMPC evaluations.
Specifically, \dpm{} fails $2$ times with \nmpcc{}, $8$ times with \rnmpc{} and $9$ times with \anmpc{}, while \ddm{} fails $4$,
$3$, and $5$ times respectively.}

 \begin{figure}[!h]
    \centering
    \includegraphics[width=1\linewidth]{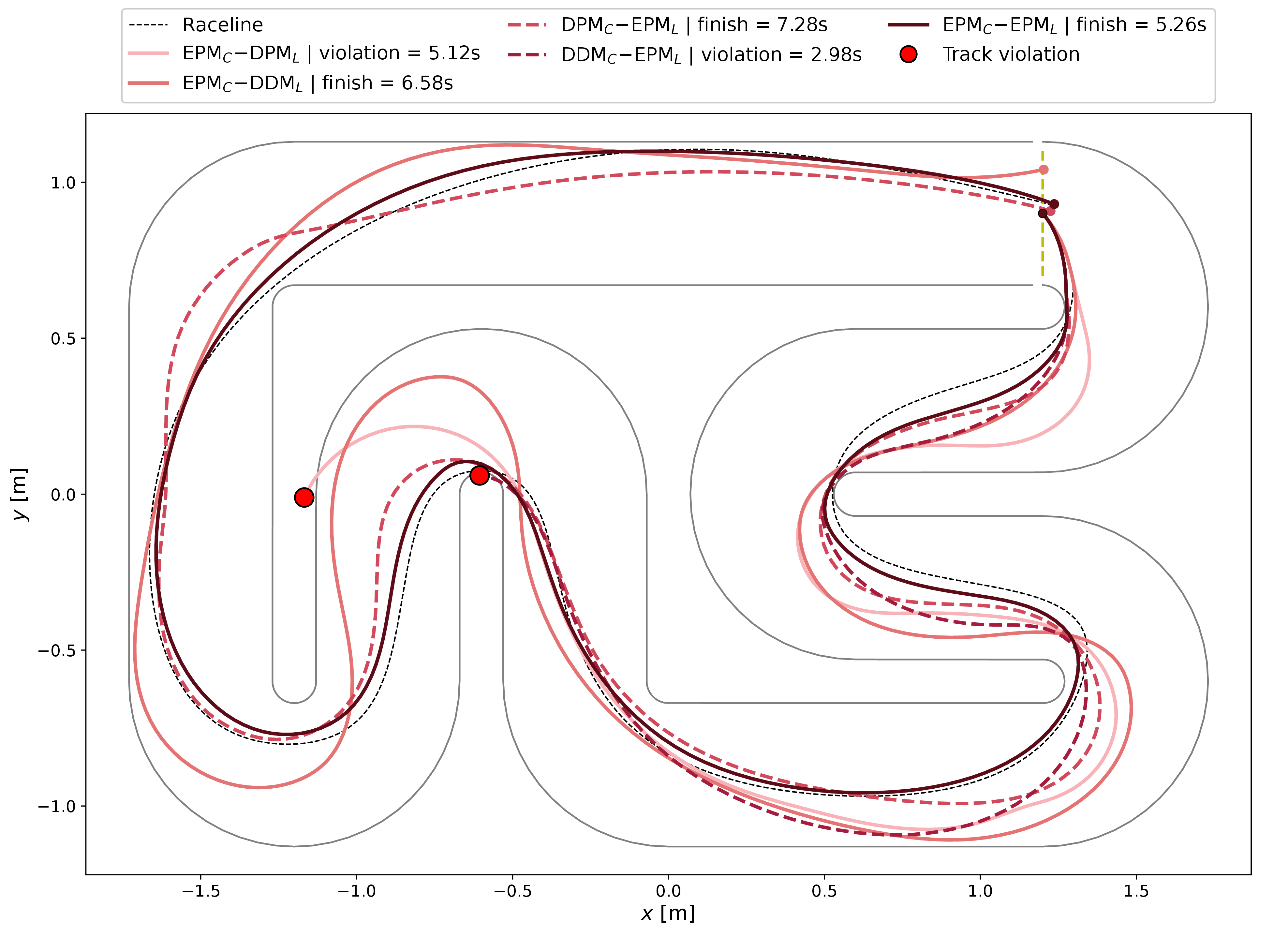}
    \caption{Trajectory comparison on ETHZMobil track under fixed internal \epm{} model with varying dynamics model (solid) and varying internal model with fixed \epm{} dynamics model (dashed).}
    \label{fig:placeholder}
\end{figure}

\bc*{Table~\ref{tab:noise_ablation} summarizes lap completion success and one lap completion time across the full uncertainty sweep.
The results show that \epm{} provides the most consistent closed-loop performance under increasing uncertainty, particularly for the NMPC-based controllers.
\nmpcc{}, \rnmpc{} and \anmpc{} paired with \epm{} complete all evaluated noise levels up to $\eta=5$.
The feedback-based PID and RISE controllers show less consistent behavior because their control is driven by instantaneous tracking, velocity, and cross-track errors, making them susceptible to the added uncertainty.
These results show that the learned load-sensitive and actuator-aware parameterization in \epm{} improves closed-loop robustness across heterogeneous controller architectures.}

\bc*{Fig.~\ref{fig:placeholder} shows the result of an MPC ablation study that isolates the effect of the internal controller \bc*{$({\cdot})_C$ model (used by the optimizer)} from the learned dynamics model $({\cdot})_L$ used for closed-loop propagation.
The solid trajectories correspond to a fixed internal \epm{} control model with three varying learned dynamic models, while dashed trajectories correspond to varying internal models and a fixed \epm{} learned dynamics model. 
The proposed EPM$_C$–EPM$_L$ configuration achieves the closest raceline adherence and fastest successful lap completion, whereas mismatch in the internal control model combinations produce larger tracking deviations and violate the track boundary, particularly in high-curvature sections.}

\section{Conclusion}
\label{sec:conclusion}
We presented CAPE, a benchmarking framework for evaluating the effect of the vehicle dynamics model on closed-loop controller performance in autonomous racing.
The framework enables comparison of heterogeneous control architectures under learned vehicle dynamics models, including a Deep Pacejka Model, a Deep-learning Dynamics Model, and an Enhanced Physics Model.
The results demonstrate that improvements in the learned vehicle dynamics can translate directly into measurable gains in closed-loop control performance.
In addition, we evaluated the performance of these learned models under realistic disturbance-aware simulation pipeline incorporating measurement noise, actuator delay, and model uncertainty. 
%
%
Future work will extend the CAPE framework to a) enable reinforcement learning control policies in the benchmarking pipeline to study how learned dynamics models influence policy training and closed-loop policy performance; and b) the framework will be validated on an experimental autonomous racing platform to evaluate how these internal vehicle dynamics models translate from simulation to real-world systems.

\section*{Acknowledgment}
ChatGPT 5.1 [OpenAI] was used  in the preparation to improve the clarity, grammar, and readability of the text.

\printbibliography
\end{document}